\documentclass[11pt,a4paper]{article}
\pdfoutput=1
\usepackage{jinstpub}         

\usepackage[colorinlistoftodos]{todonotes}

\usepackage{epsfig}
\usepackage{amssymb}
\usepackage{subfigure}
\usepackage{lineno}

\usepackage{titlesec}
\titleformat{\paragraph}[runin]
{\bfseries\scshape}{\theparagraph}{1em}{}
\setcitestyle{square}

\input{symbols.tex}

\date{\today}

\title{The Simulation of the Sensitivity of the Antarctic Impulsive Transient Antenna (ANITA)
      to Askaryan Radiation from Cosmogenic Neutrinos Interacting in the Antarctic Ice}

\collaboration{ANITA Collaboration}

\affiliation[a]{Dept. of Physics and Astronomy, University College London, \\ Gower Street, London, United Kingdom.}
\affiliation[b]{Dept. of Physics, Center for Cosmology and AstroParticle Physics, Ohio State Univ., \\ W Woodruff Avenue, Columbus, OH 43210, U.S.A.}
\affiliation[c]{Dept. of Physics, Enrico Fermi Inst., Kavli Inst. for Cosmological Physics, Univ. of Chicago, \\ S Ellis Avenue, Chicago, IL 60637, U.S.A.}
\affiliation[d]{LSST, \\ North Cherry Avenue, Tucson, AZ 85721, U.S.A.}
\affiliation[e]{Jet Propulsion Laboratory, \\ Oak Grove Drive, Pasadena, CA 91109, U.S.A.}
\affiliation[f]{Dept. of Physics and Astronomy, Univ. of Kansas, \\ Wescoe Hall Drive, Lawrence, KS 66045, U.S.A.}
\affiliation[g]{National Research Nuclear University, Moscow Engineering Physics Institute, \\ Kashirskoye Highway, Russia 115409, Russia.}
\affiliation[h]{Dept. of Physics and McDonnell Center for the Space Sciences, Washington Univ. in St. Louis, \\ Brookings Drive, St. Luis, MO 63130, U.S.A.}
\affiliation[i]{Dept. of Physics, Univ. of Delaware, \\ The Green, Newark, DE 19716, U.S.A.}
\affiliation[j]{Dept. of Physics, Grad. Inst. of Astrophys.,\& Leung Center for Cosmology and Particle Astrophysics, National Taiwan University, \\ Roosevelt Road, 10617 Taipei, Taiwan.}
\affiliation[k]{Dept. of Physics and Astronomy, Univ. of Hawaii, \\ Correa Road, Manoa, HI 96822, U.S.A.}
\affiliation[l]{Astrophysical Institute, Vrije Universiteit Brussel, \\ Pleinlaan 2, 1050, Brussels, Belgium.}
\affiliation[m]{Center for Astrophysics and Space Sciences, Univ. of California, San Diego,  \\ Gilman Dive, La Jolla, CA 92093, U.S.A.}
\affiliation[n]{Dept. of Physics and Astronomy, Univ. of California, Los Angeles, \\ Portola Plaza, Los Angeles, CA 90095, U.S.A.}
\affiliation[o]{Max-Planck-Institut f\"{u}r Kernphysik, \\ Saupfercheckweg, 69117 Heidelberg, Germany.}
\affiliation[p]{Physics Dept., California Polytechnic State Univ., \\ North Perimeter Road, San Luis Obispo, CA 93407, U.S.A.}

\author[a,1]{L.~Cremonesi\note{Corresponding Author},}\emailAdd{l.cremonesi@ucl.ac.uk}
\author[b]{A.~Connolly,} 
\author[b]{P.~Allison,}
\author[b]{O.~Banerjee,}
\author[a]{L.~Batten,} 
\author[b]{J.~J.~Beatty,} 
\author[c,d]{K.~Bechtol,}
\author[e]{K.~Belov,}
\author[f,g]{D.~Z.~Besson,}
\author[h]{W.~R.~Binns,} 
\author[h]{V.~Bugaev,} 
\author[i]{P.~Cao,} 
\author[j]{C.~C.~Chen,} 
\author[j]{C.~H.~Chen,}
\author[j]{P.~Chen,} 
\author[i]{J.~M.~Clem,} 
\author[b]{B.~Dailey,} 
\author[c]{C.~Deaconu,} 
\author[h]{P.~F.~Dowkontt,} 
\author[k]{B.~D.~Fox,} 
\author[b]{J.~W.~H.~Gordon,}
\author[k]{P.~W.~Gorham,} 
\author[k]{B.~Hill,} 
\author[j]{J.~J.~Huang,}
\author[b,c]{K.~Hughes,}
\author[b]{R.~Hupe,} 
\author[h]{M.~H.~Israel,} 
\author[e]{K.~M.~Liewer,}
\author[j]{S.~Y.~Lin,}
\author[j]{T.~C.~Liu,} 
\author[c]{A.~B.~Ludwig,} 
\author[k]{L.~Macchiarulo,} 
\author[k]{S.~Matsuno,} 
\author[b]{K.~McBride,}
\author[k]{C.~Miki,} 
\author[i,l]{K.~Mulrey,}
\author[j]{J.~Nam,}
\author[a]{R.~J.~Nichol,}
\author[f,g]{A.~Novikov,} 
\author[c]{E.~Oberla,} 
\author[b]{S.~Prohira,}
\author[h]{B.~F.~Rauch,}
\author[k,m]{J.~M.~Roberts,}
\author[e]{A.~Romero-Wolf,}
\author[k]{B.~Rotter,}
\author[k]{J.~W.~Russell,} 
\author[n]{D.~Saltzberg,}
\author[i]{D.~Seckel,}
\author[k,o]{H.~Schoorlemmer,}
\author[j]{J.~Shiao,}
\author[b]{S.~Stafford,}
\author[f]{J.~Stockham,}
\author[f]{M.~Stockham,}
\author[n]{B.~Strutt,} 
\author[p]{J.~Stuhr,}
\author[b]{M.~Sutherland,}
\author[k]{G.~S.~Varner,}
\author[c]{A.~G.~Vieregg,}
\author[j]{S.~H.~Wang,}
\author[p]{S.~A.~Wissel}

\abstract{A Monte Carlo simulation program for the radio
      detection of Ultra High Energy (UHE) neutrino interactions in
      the Antarctic ice as viewed by the Antarctic Impulsive Transient Antenna (ANITA) is described in this article. 
      The program, \icemc,  provides an input spectrum of UHE neutrinos, the
      parametrization of the Askaryan radiation generated by their
      interaction in the ice, and the propagation of the radiation
      through ice and air to a simulated model of the third and fourth
      ANITA flights.
      This paper provides an overview of the \icemc simulation, 
      descriptions of the physics models used and of
      the ANITA electronics processing chain, data/simulation
      comparisons to validate the predicted performance, and a summary
      of the impact of published results.}


\arxivnumber{1903.11043}

\begin{document}

\maketitle

\flushbottom

\section{Introduction}
The ANtarctic Impulsive Transient Antenna (ANITA) experiment is a NASA-funded, balloon-borne experiment aiming primarily to detect Ultra-High-Energy (UHE) neutrinos and cosmic rays~\cite{ANITA1detector,ANITA1paper,ANITA2paper,ANITA2erratum}.
UHE neutrinos may be produced when cosmic rays interact with the
cosmic microwave background, either through the Greisen Zatsepin
Kuzmin (GZK)~\cite{greisen1966end,zatsepin1966gt} mechanism or through photo-disintegration, or directly from an astrophysical source.
UHE neutrinos interact in the Antarctic ice and produce a coherent radio pulse of Askaryan radiation~\cite{askaryan}. UHE cosmic rays can also interact with the geomagnetic field to produce a coherent radio impulse~\cite{suprun2003synchrotron,falcke2003detecting,ANITA1UHECR}. The geometry of the Earth's magnetic field in Antarctica allows ANITA to distinguish the signatures of these two types of events through their polarization.

\begin{figure}[!h]\centering
  \includegraphics[width=.8\linewidth]{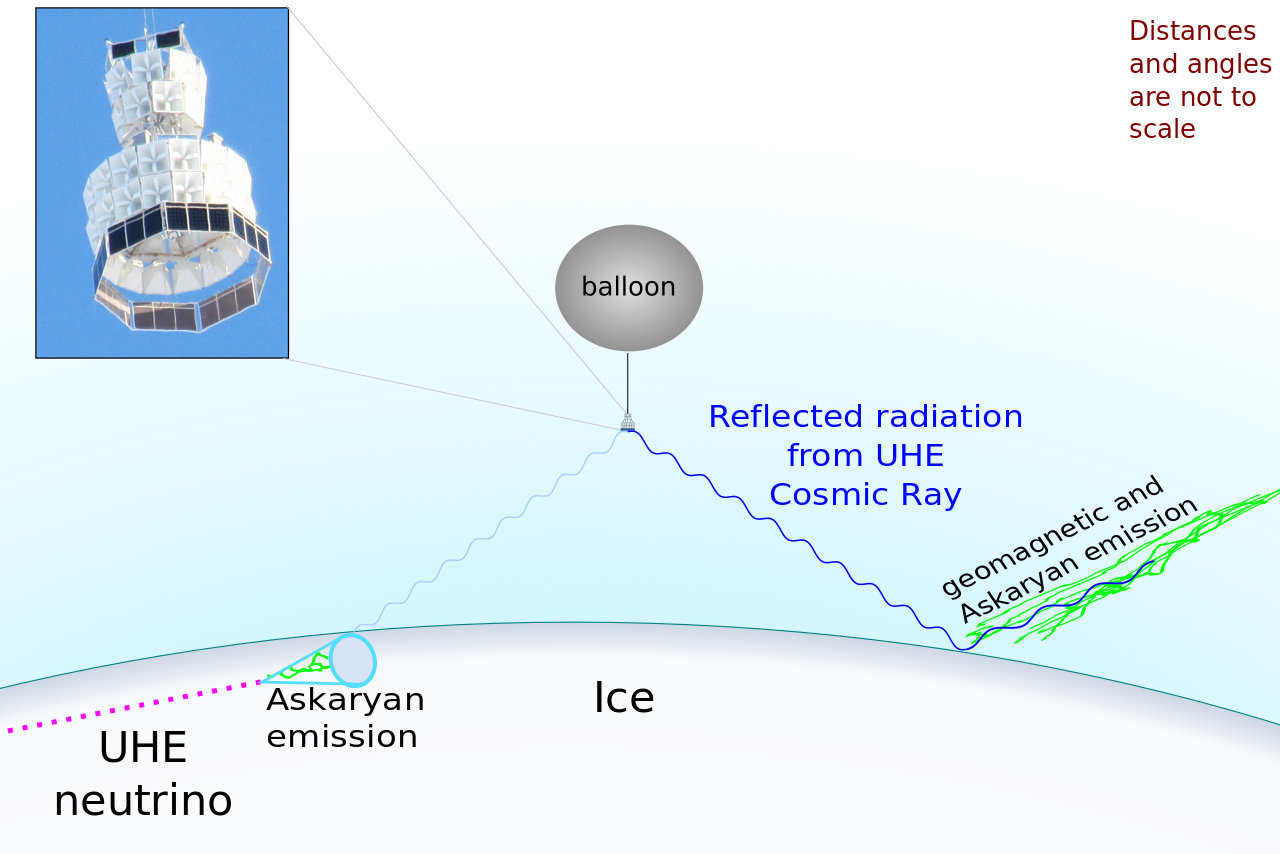}
  \caption{The ANITA detection concept with a photo of the ANITA-IV
    payload. UHE neutrinos interact with the Antarctic ice and produce
    a coherent radio pulse of Askaryan radiation. UHE cosmic ray
    interactions in the atmosphere produce a shower of secondary
    particles that interact with the geomagnetic field and can also
    produce a coherent radio impulse. Both these signals are detected by the ANITA instrument. 
  }
  \label{fig:intro_ANITAconcept}
\end{figure}

The $\sim$8\,m tall ANITA experiment flies about 40\,km above Antarctica looking for radio signals in the band between 200 and 1200\,MHz, produced by UHE neutrinos and cosmic rays.
Figure~\ref{fig:intro_ANITAconcept} shows a basic scheme of the
Askaryan neutrino detection at the ANITA experiment, as well as the UHE cosmic ray detection principle.
Four ANITA flights (Section~\ref{sec:anita3}) have been successfully
completed as of 2019, and this paper focuses on the simulations of the third and fourth ANITA flights.

The \icemc program is a C++ Monte Carlo simulation program based on
ROOT~\cite{brun1997root} used to simulate the Askaryan radiation produced
by neutrino interactions and the response of the ANITA detector to this
radiation.
This program is used by the ANITA collaboration to tune the selection
cuts of the cosmogenic neutrino analysis and quantify the experiment's sensitivity.

\begin{figure}[!h]\centering
  \includegraphics[width=.8\linewidth]{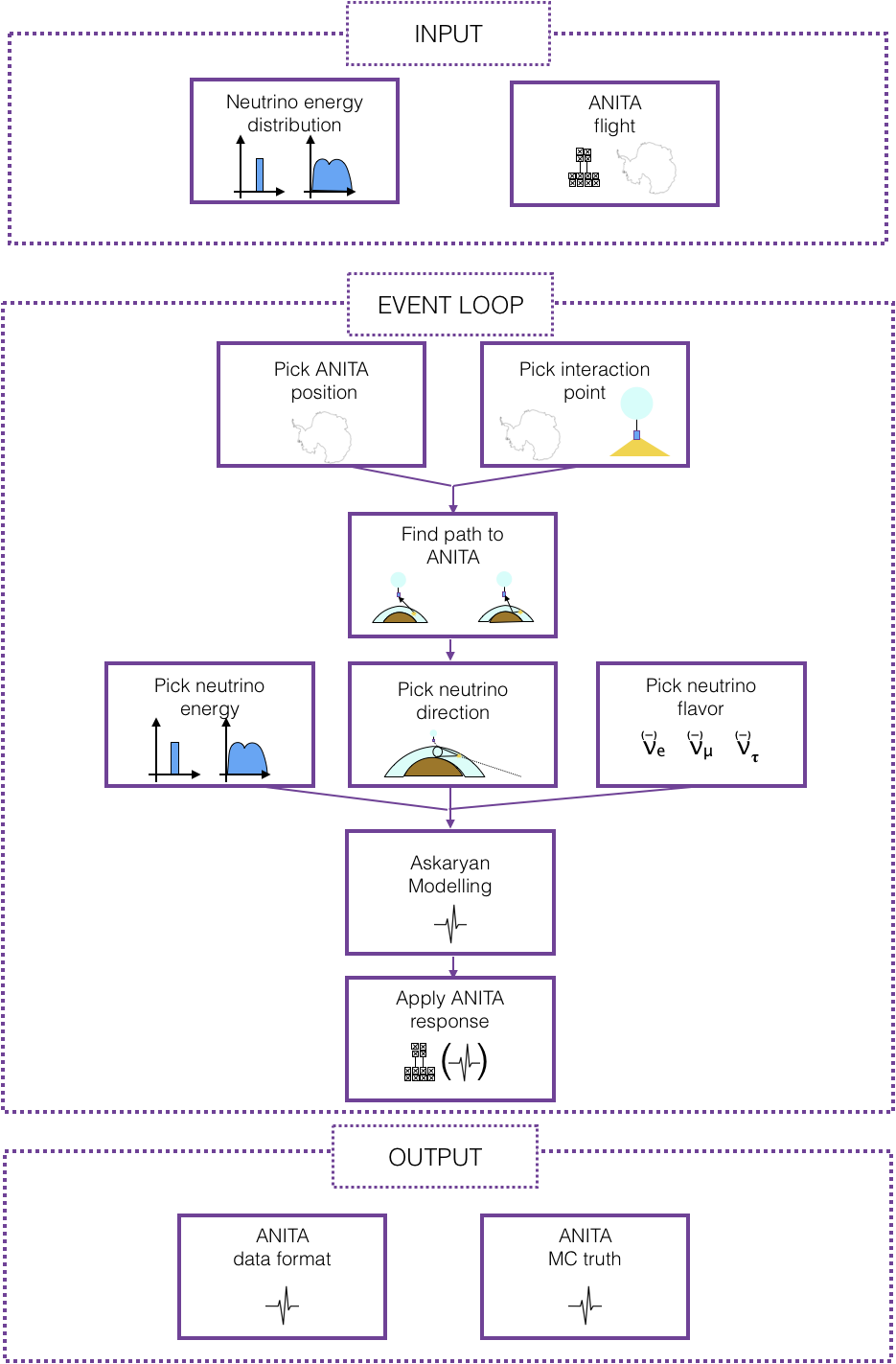}
  \caption{Flowchart of the \icemc simulation for a single candidate.}
  \label{fig:intro_icemcFlow}
\end{figure}

A flowchart of the \icemc simulation steps is shown in Figure~\ref{fig:intro_icemcFlow}.
At the beginning of each run the user can choose which ANITA flight to simulate and which neutrino energy spectrum to use.
For each neutrino event, the position of the payload along the ANITA flight path, as well as the neutrino interaction position are randomly chosen.
These are used to find the path to the ANITA detector (Section~\ref{sec:eventGeometry}). 
The neutrino direction is then chosen within detectable angles; the neutrino energy is chosen following the input theoretical model; and the neutrino flavor and interaction type are chosen according to their expected ratios. 
These parameters are used to produce the radio-frequency pulse following the Askaryan model described in Section~\ref{sec:rf}.
The signal is then propagated through the ice and air to the ANITA
detector (see Section~\ref{sec:propagation}). 
The response of the ANITA signal chain is simulated, and the resulting 
data are saved in the same format as ANITA flight data 
(see Section~\ref{sec:ANITA}).
Different parts of the simulation are validated against data taken
both at the NASA Long Duration Balloon Facility
before the launch and in-flight during the past ANITA flights (see
Section~\ref{sec:validation}).
Finally the neutrino acceptance of the third and fourth ANITA flights, and
the contributions of the different parts of the simulation to the uncertainties on the acceptance are presented in Section~\ref{sec:results} .

\section{The ANITA flights}
\label{sec:anita3}
The first two ANITA payloads flew in 2006-2007\cite{ANITA1paper} and 2008-2009\cite{ANITA2paper,ANITA2erratum}, respectively.
Although \icemc can be used to simulate older flights, this paper focuses on the simulation of the third and fourth flights.

\begin{figure}[!h]\centering
  \includegraphics[width=.45\linewidth]{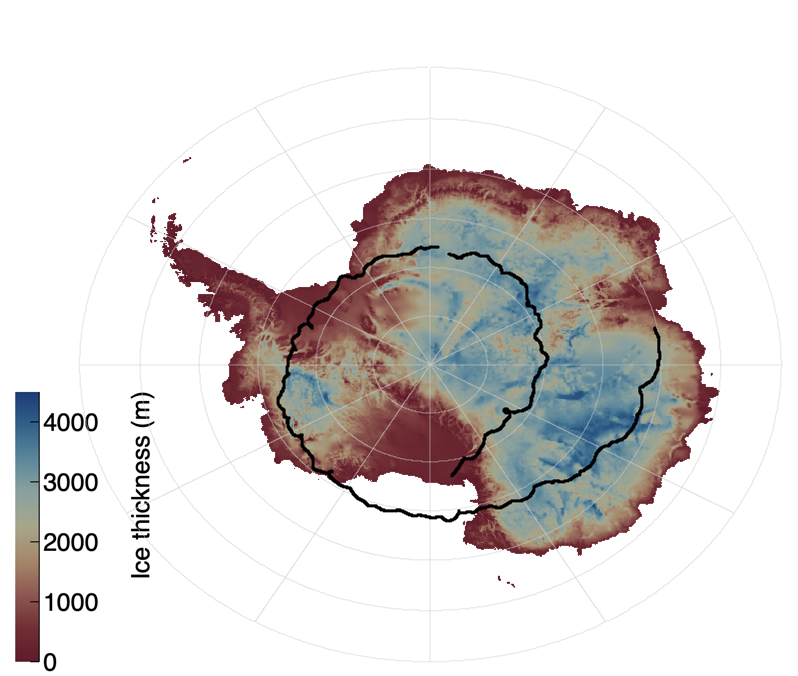}
  \includegraphics[width=.45\linewidth]{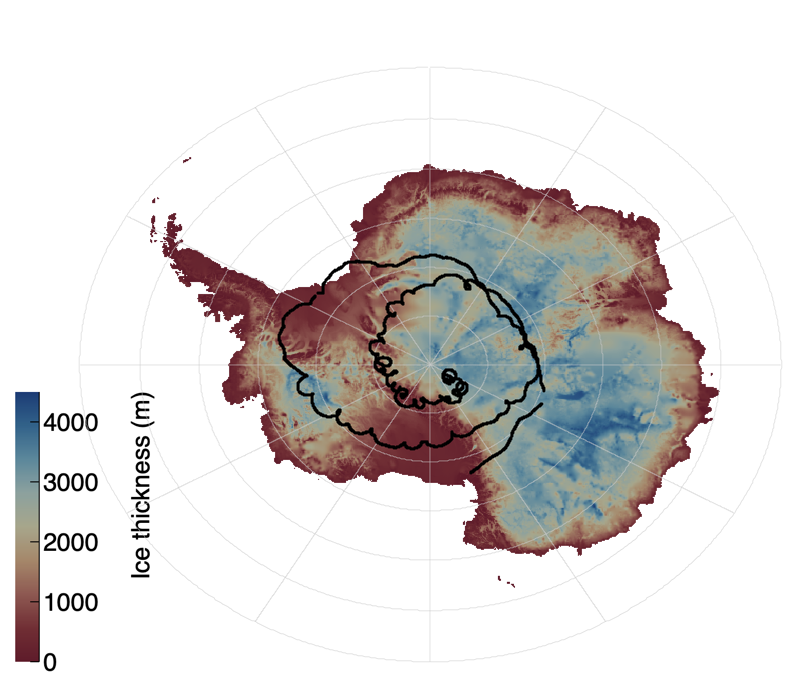}
  \caption{Flight path simulated in \icemc for the ANITA-III (left) and ANITA-IV (right) flights.
    Antarctica map produced using the BEDMAP model~\cite{bedmap}.     }
  \label{fig:ANITA_flightPath}
\end{figure}

Figure~\ref{fig:ANITA_flightPath} shows the flight paths of the ANITA-III and ANITA-IV instruments. 
The color map shows the Antarctic ice depth. 
The third ANITA payload (ANITA-III) launched on December 18$^{\text{th}}$, 2014 from the
NASA Long Duration Balloon (LDB) facility near McMurdo Station, Antarctica.
ANITA-III followed the polar vortex flying at the altitude of 37\,km for
22 days until January 9$^{\text{th}}$, 2015 when the flight was terminated
near the Australian Davis Station.
Similarly, the fourth ANITA payload (ANITA-IV) launched on December 2$^{\text{nd}}$, 2016 from the NASA LDB facility in Antarctica and landed approximately 100\,km from the South Pole Station on December 29$^{\text{th}}$, 2016.

The ANITA-III and ANITA-IV instruments were similar to the two previous
ones~\cite{ANITA1paper,ANITA2paper}, using 48 dual-polarization quad-ridged horn antennas with bandwidth from 200 to 1200\,MHz. 
The antennas were arranged in three rings forming 16 azimuthal sectors per ring, with the top ring divided into two layers to fit the launch envelope specifications, as shown in the top left of Figure~\ref{fig:intro_ANITAconcept}.


As the payload rotates freely, two sets of GPS units were used
independently to determine the
payload position and attitude.
Power was supplied by an octagonal array of photovoltaic panels and stored using four pairs of 12-V lead-acid batteries.
Communication to and from the payload was through the Iridium and TDRSS satellite systems throughout the flight, and by a direct line-of-sight radio link when within range of McMurdo Station.

\section{Event geometry}
\label{sec:eventGeometry}

\subsection{Modeling the Antarctic continent}
Crust 2.0~\cite{crust2} is used to model the Earth's interior near the 
surface.  It is based on seismological data published from the Cooperative Studies of the Earth's Deep Interior (CSEDI).
The model gives thicknesses and densities of seven material
layers in $2^{\circ} \times 2^{\circ}$ bins:  ice, water, soft sediments, hard sediments, upper crust,
middle crust, and lower crust.  

The total Antarctic ice volume is computed by summing the product of ice
thickness and surface area for each bin within the Antarctic continent.
It is also possible to run the simulation using BEDMAP ice thickness
and subglacial topographic model of Antarctica, developed by the
British Antarctic Survey~\cite{bedmap}.
This model has higher resolution, 
but it is much slower to run, so by default we use
Crust 2.0.
Using Crust 2.0, the \icemc program finds $2.976 \times 10^{16}$ m$^3$ of Antarctic ice in this model, compared to $3.011 \times 10^{16}$ m$^3$ reported by the US Geological Survey~\cite{usgs}, a 1.15\% difference.

\subsection{Picking neutrino interaction point and direction}
\label{sec:pickneutrino}

For each neutrino event, the payload position is chosen at random from a set of positions along the ANITA flight path (see Figure~\ref{fig:ANITA_flightPath}).
To simulate only those neutrino interactions that might lead to a detectable signal, interaction positions are limited to occur within the horizon as seen by the payload (roughly between 700 and 800\,km from the payload), and neutrino directions are chosen from an annulus on the sky consistent with viewing angles detectable at the payload (see Section~\ref{sec:weights}). 
The maximum angle that the ray may diverge from the axis of the Cherenkov cone for the interaction to still be detectable depends on the electric field, the distance from the interaction to the payload, and shower nature: this is typically between 10 and 13 degrees.

Both direct and reflected detection are simulated.
In the first case the signal is propagated from the interaction position upwards to the payload. In the second case, the signal is propagated downwards
towards to the ice-rock interface, approximated as a flat mirror,
where the signal is then reflected upwards towards the payload.

\subsection{Event weighting}
\label{sec:weights}
To reduce computation time,
each event is weighted according to the probability of the neutrino reaching the interaction point without being absorbed in the earth, as well as
the "phase space" reduction (defined below), such that only those topologies that give measurable signal are fully simulated. 
As a neutrino moves through the earth, it encounters varying
densities as it passes through layers of the earth's interior,
and thus differing interaction lengths. 
The neutrino survival probability is calculated from the along-track
water-equivalent amount of material traversed.

\begin{figure}[!h]\centering
  \includegraphics[width=.4\linewidth, trim = 0 6.5cm 14cm 0, clip]{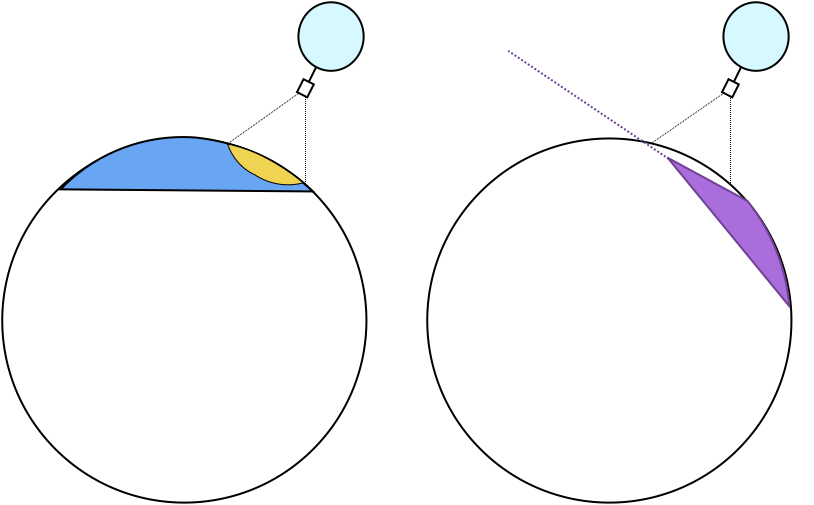}
  \caption{A schematic of the position weight used in \icemc. The
    position weight arises because only interaction positions within
    the balloon horizon are simulated (yellow over blue area, where
    the blue area represents the entire Antarctica).
 }
  \label{fig:weights}
\end{figure}

The phase space factor is the product of the weights derived from the neutrino interaction position and neutrino direction.
These weights are assumed to be independent of one another.
The position weight arises due to the neutrino interaction position being
chosen only within the payload horizon.
This is calculated as
the ratio of the volume of ice within the horizon for the $i^{th}$
event and the total volume of ice in Antarctica (ratio of the yellow to blue
volumes in Figure~\ref{fig:weights}).
The neutrino direction is chosen such that the axis of the Cherenkov cone
lies close in angle to the direction from the interaction point to the payload. 
The direction weight is calculated as the ratio of the solid angle coming from the Cherenkov cone and a unit sphere.

\section{Radio signal simulation} 
\label{sec:rf}

\subsection{Askaryan parametrization}
For each event the neutrino energy is picked randomly following the input energy spectrum.
Rather than simulating the particle shower development in the ice that
would produce the Askaryan radiation, a 1D parametrization is used 
to find the peak of the Askaryan signal at 1\,m from the point of interaction, $\mathcal{E}^{(\mathrm{@ 1m})}$, 
according to Reference~\cite{JaimeAskarian2000} and following:
\begin{equation}
\label{eq:vmmhz}
\mathcal{E}^{(\mathrm{@ 1m})} =2.53\times 10^{-7}\cdot \frac{\sin{\theta_{\mathrm{view}}}}{\sin{\theta_{\mathrm{Ch}}}}\cdot  \frac{E}{\mathrm{TeV}} \cdot \frac{\nu}{\nu_0} \cdot \frac{1}{1+\left( \frac{\nu}{\nu_0} \right)^{1.44}} \;,
\end{equation}
\noindent where $\nu$ is the frequency, $\nu_0=1.15$\,GHz, $E$ is
the shower energy (see Subsection~\ref{subsec:emhadshower}),
$\theta_{\mathrm{view}}$ is the viewing angle and $\theta_{\mathrm{Ch}}$ is the Cherenkov angle.
This parametrization is valid up to 5\,GHz, covering the ANITA
band of 0.2-1.2\,GHz.

\subsection{Neutrino flavors and interaction types}
\label{subsec:emhadshower}
The \icemc program assumes flavor democracy, assuming 
the flavors are fully mixed before the neutrinos get to the ice.
Information about each flavor is stored separately so the sensitivity to each flavor may be quoted separately.

The neutrino undergoes a charged/neutral current interaction in
$68.7\%/31.3\%$ of cases.
For charged current electron neutrino interactions, the electromagnetic
component is calculated as $1-y$, where $y$ is the inelasticity and is
approximated by a double-exponential function~\cite{gandhi}. 
In all other cases the electromagnetic component is considered negligible.
The hadronic component is equal to the inelasticity.
Secondary particles are not propagated.

\subsection{Cherenkov Cone}
\label{subsec:cherenkov_width}
The width of the Cherenkov cone is parametrized separately for the electromagnetic and hadronic components of the shower, according to References \cite{JaimeAskarian2000} and \cite{jaime05}, respectively.
The width of the electromagnetic
component in degrees is characterized by:
\begin{equation}
\label{eq:deltheta_em}
\Delta\theta_{em}(\nu)=2.7^{\circ} \cdot \frac{\nu_0}{\nu}\cdot \left(
  \frac{E_{LPM}}{ 0.14 E_{EM}+E_{LPM}} \right)^{0.3} \;,
\end{equation}
where 
$E_{EM}$ is the part of the shower energy associated with electromagnetic particles, and 
$E_{LPM}$ is the energy above which the Landau-Pomeranchuk-Migdal
(LPM) effect becomes important.  
The LPM effect causes
the bremsstrahlung interaction to become suppressed when the longitudinal
momentum transfer ($q \propto k/E^2$, where $k$ is the photon energy and $E$ is the electron energy) becomes very small. In this case, the 
Heisenberg uncertainty
causes the interaction to occur over many
scattering centers, resulting in destructive interference.  
This effect reduces the width of the Cherenkov cone, but not the magnitude of the electric field at the Cherenkov angle.
Following the recommendation in Reference~\cite{JaimeAskarian2000}, 
$E_{LPM}$ is set to $2\times10^{15}\ev$ in ice,
and can be scaled to other media using the ratio of the respective radiation lengths.

The width of the Cherenkov cone for hadronic showers is modeled as laid out in Equation 9
in~\cite{jaime05}:
\begin{equation}
\Delta \theta_{\mathrm{had}} (\nu) =\frac{c}{\nu} \cdot
	\frac{\rho}{\mathrm{K}_{\Delta} X_0} \cdot
	\frac{1}{n^2-1} \;,
\end{equation}
\noindent where
$c$ is the speed of light,
$\nu$ is the frequency considered, 
$\rho$ is the density of ice,
$\mathrm{K}_{\Delta}$ is a normalization constant determined by a
separate Monte Carlo simulation~\cite{jaime05},
$X_0$ is the radiation length, and
$n$ is the index of refraction.

The signal strength at viewing angle $\theta_{\mathrm{view}}$ 
away from the Cherenkov angle $\theta_{\mathrm{Ch}}$ is also parametrized following
Equation\,13 in Reference\,~\cite{jaime05}:
\begin{equation}
\mathcal{E}(\theta_{\mathrm{view}})=\frac{\sin{\theta_{\mathrm{view}}}}{\sin{\theta_{\mathrm{Ch}}}} \cdot
\mathcal{E}(\theta_{\mathrm{Ch}})\cdot \exp\left[-\left(
    \frac{\theta_{\mathrm{view}}-\theta_{\mathrm{Ch}}}{\Delta\theta_{\mathrm{em,had}}} \right)^2
\right] \;.
\end{equation}

\section{Propagation in ice and air}
\label{sec:propagation}

The electric field is propagated to the payload using a standard ray tracing algorithm.
The electric field magnitude at the payload position (before applying
the ANITA instrument response) is calculated as follows:
\begin{equation}
 \mathcal{E}_{\perp,||} = \mathcal{E}^{(\mathrm{@ 1m})}  
 \mathrm{e}^{(-d_{\mathrm{ice}}\ /\ \ell_{\mathrm{attn}})}
 \frac{t_{\perp,||}}{d_{\mathrm{ice}} + d_{\mathrm{air}}}
 \ F(\theta_{\mathrm{view}}-\theta_{\mathrm{Ch}}) \,.
\end{equation}
The first factor after $\mathcal{E}^{(\mathrm{@ 1m})}$ accounts for propagation in ice, with $d_{\mathrm{ice}}$ being the path length and $\ell_{\mathrm{attn}}$ the attenuation length in ice. 
The second factor accounts for the refraction from ice to air, using
the specular Fresnel coefficients, $t_{||}$ and $t_{\perp}$, for the parallel and perpendicular components of the normalized transmitted electric field vector at the ice-air interface with respect to the local surface normal.
The surface normal includes a Gaussian 1.2\% direction perturbation to account for surface slope effects.
Finally, $F(\theta_{\mathrm{view}}-\theta_{\mathrm{Ch}})$ is a geometrical attenuation factor of the field strength in air resulting from viewing the Cherenkov emission at an angle $\theta_{\mathrm{view}}$ different from the Cherenkov angle $\theta_{\mathrm{Ch}}$.
The functional form of $F$ is taken to be Gaussian and is set to zero for ($\theta_{\mathrm{view}}-\theta_{\mathrm{Ch}}$) > 20 $\Delta \theta_{\mathrm{Ch}}$, where $\Delta \theta_{\mathrm{Ch}}$ is the width of the Cherenkov cone.

Cherenkov radiation is radially polarized, and the in-ice polarization vector of the Cherenkov radiation is radially outward in the plane formed by
the neutrino velocity vector and the Cherenkov propagation vector.
Attenuation lengths for radio in Antarctica are based on measurements performed at the Ross Ice Shelf and the South Pole~\cite{barwick2005south,barrella2011ross}.
The index of refraction is taken as 1.79 for deep ice and 1.325 at the surface. A model for the firn, a layer of packed snow above the ice, based on data taken by the RICE Collaboration~\cite{PhysRevD.73.082002} is used at depths shallower than 150\,m.

Surface roughness acts to allow different portions of the ice surface to contribute transmitted power to the payload for a given event, because it allows new scattering geometries at the air-ice interface.
The University of Hawaii ANITA group developed a simulation program, used for the ANITA-I and ANITA-II flights, which included surface roughness effects. They found that ice surface roughness contributed to an increase in acceptance of roughly a factor of 50\% at low energy (close to $10^{18}$\,eV) and 40\% at higher energy.
\icemc currently does not include ice surface roughness effects, and hence provides a conservative estimate of the sensitivity of the ANITA flights. Future versions of \icemc will include these effects.


\section{ANITA detector model}
\label{sec:ANITA}
The simulated signal is propagated to the front of each of the ANITA antennas and through the
trigger and digitizer paths of the ANITA instrument, taking into
account the payload rotation.
The payload position chosen along the flight path is used to load appropriate information about the payload rotation, channel threshold values and channel masking.
The payload geometry is simulated using photogrammetry measurements
and phase-center calibration measurements taken with a ground pulser
during the flights.
Antenna gains measured prior to the flight are
applied to the signal according to the incident angles on the E and H planes.

\begin{figure}[!h]\centering
  \includegraphics[width=.95\linewidth]{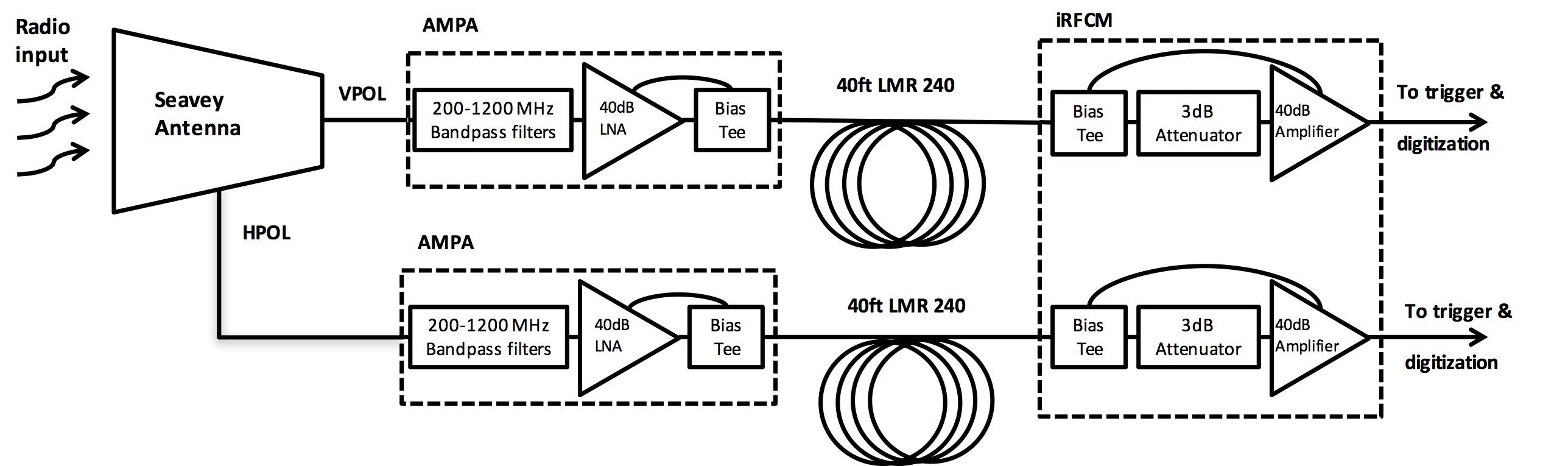}
    \includegraphics[width=.95\linewidth]{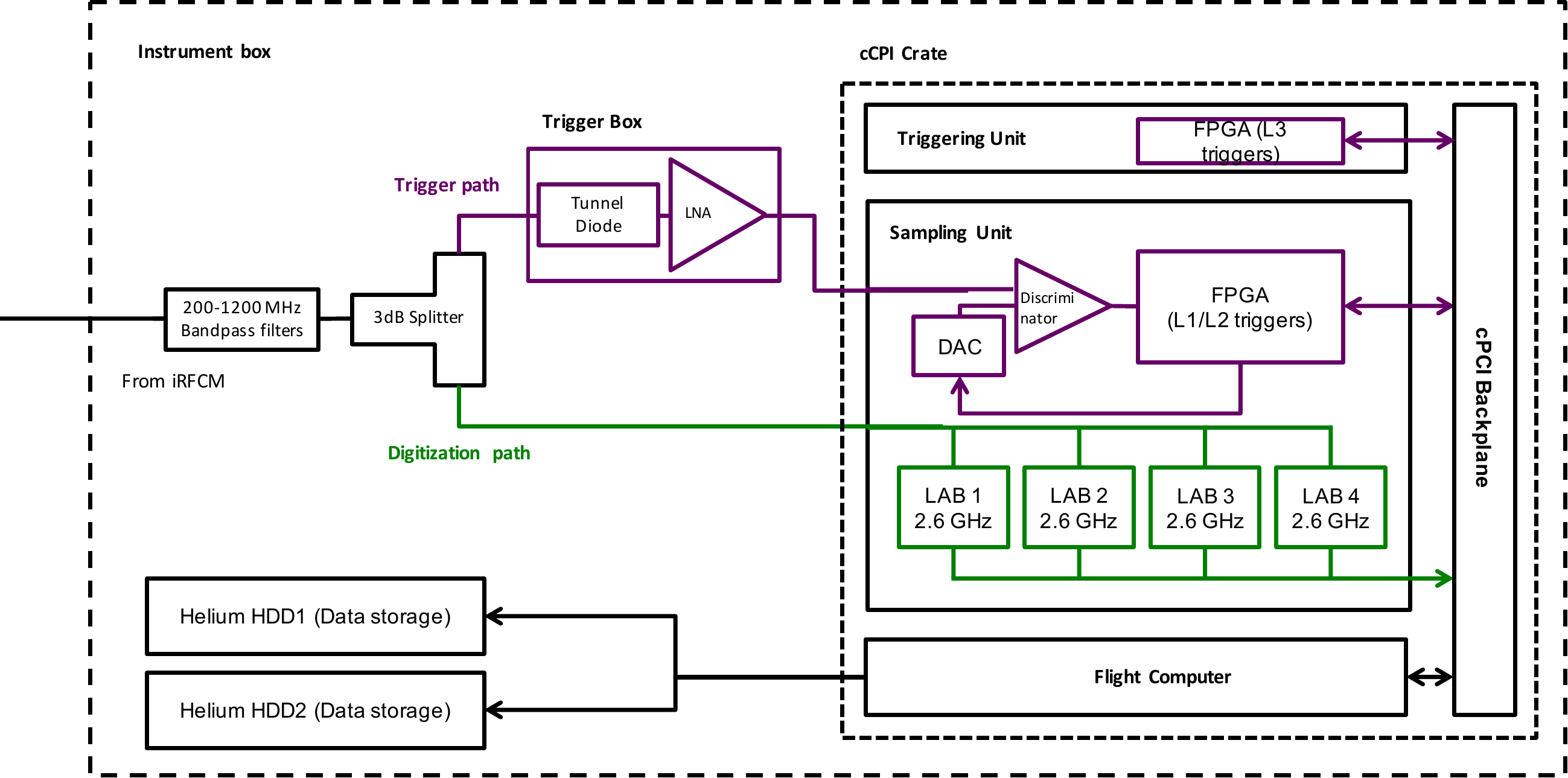}
  \caption{ANITA-III signal chain.
  The receiving antenna splits signals into vertical and horizontal
  polarizations which follow identical paths. 
After going through filters and amplifiers, the signal is split into
the trigger and digitizer paths. In the trigger path the signal passes
through a tunnel diode and is compared to a threshold. 
When a trigger is issued, a switched capacitor array digitizes the
signal with 260 samples per channel and a mean sample rate of 2.6\,GSa/s. 
The triggering and sampling units, as well as the single board flight computer are contained in the Compact Peripheral Component Interface (cPCI) crate.
For ANITA-IV, the TUFFs replace the iRFCMs, performing the second stage amplification and filtering.}
  \label{fig:ANITA3_signalChain}
\end{figure}

Figure~\ref{fig:ANITA3_signalChain} shows a schematic of the ANITA-III signal chain. 
The receiving antenna measures vertical (V-POL) and horizontal (H-POL) polarization components of incident radio waves, which then follow twin paths. 
After passing through filters and amplifiers, the signal is split into
the trigger and digitizer paths.
In the trigger path the signal passes
through a tunnel diode which acts as a square-law detector over the
input. The output of the tunnel diode is compared to the channel
threshold, which is dynamically adjusted by a PID loop to maintain the
rate for each antenna around 500\,kHz. 
ANITA-IV used $90^{\circ}$ hybrids to convert the signals into left- and right- circularly polarized (LCP and RCP) components. 
The ANITA-IV trigger logic also required a coincidence between LCP and RCP signals within a 4\,ns coincidence window to select for linearly polarized signals.
A combination of first-level (L1) and second-level (L2) triggers forms a
global trigger which can be issued in each polarization.
When a trigger is issued, a switched capacitor array digitizes the RF signal with 260 samples per channel and a mean sample rate of 2.6\,GSa/s. 
To avoid continuous wave (CW) noise from satellite and station transmissions, the ANITA-IV signal chain included the Tunable Universal Filter Front-end boards (see Subsection~\ref{subsec:tuffs} and Reference~\cite{Allison:2017vtk}), placed instead of the internal Radio Frequency Control Modules (iRFCMs). 
In the case of the ANITA-III payload a channel is the V-POL or H-POL component of one antenna; for ANITA-IV a trigger channel is the LCP or RCP component, and a digitized channel refers to the V-POL or H-POL component of one antenna.

\begin{figure}[!h]\centering
  \includegraphics[width=.45\linewidth]{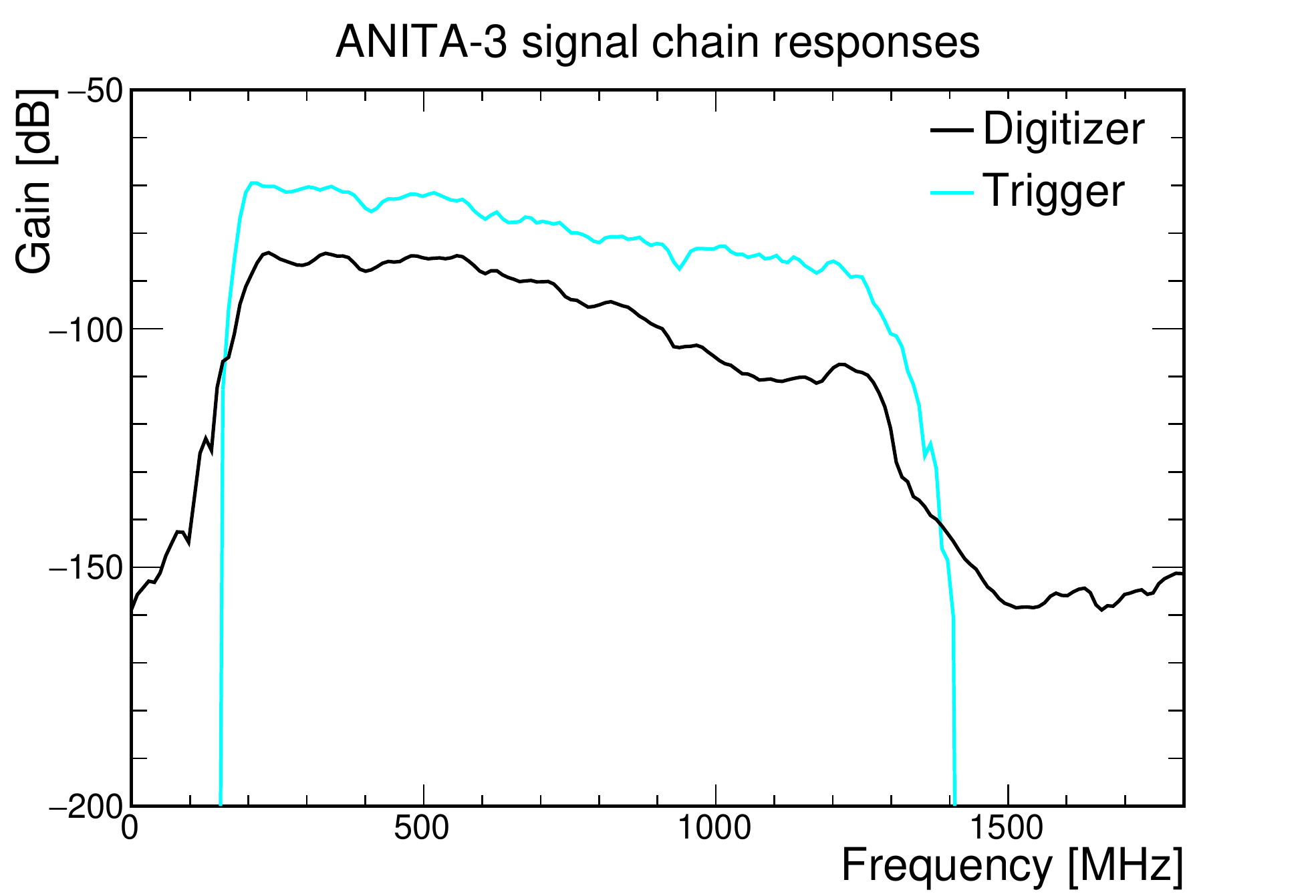}
  \includegraphics[width=.45\linewidth]{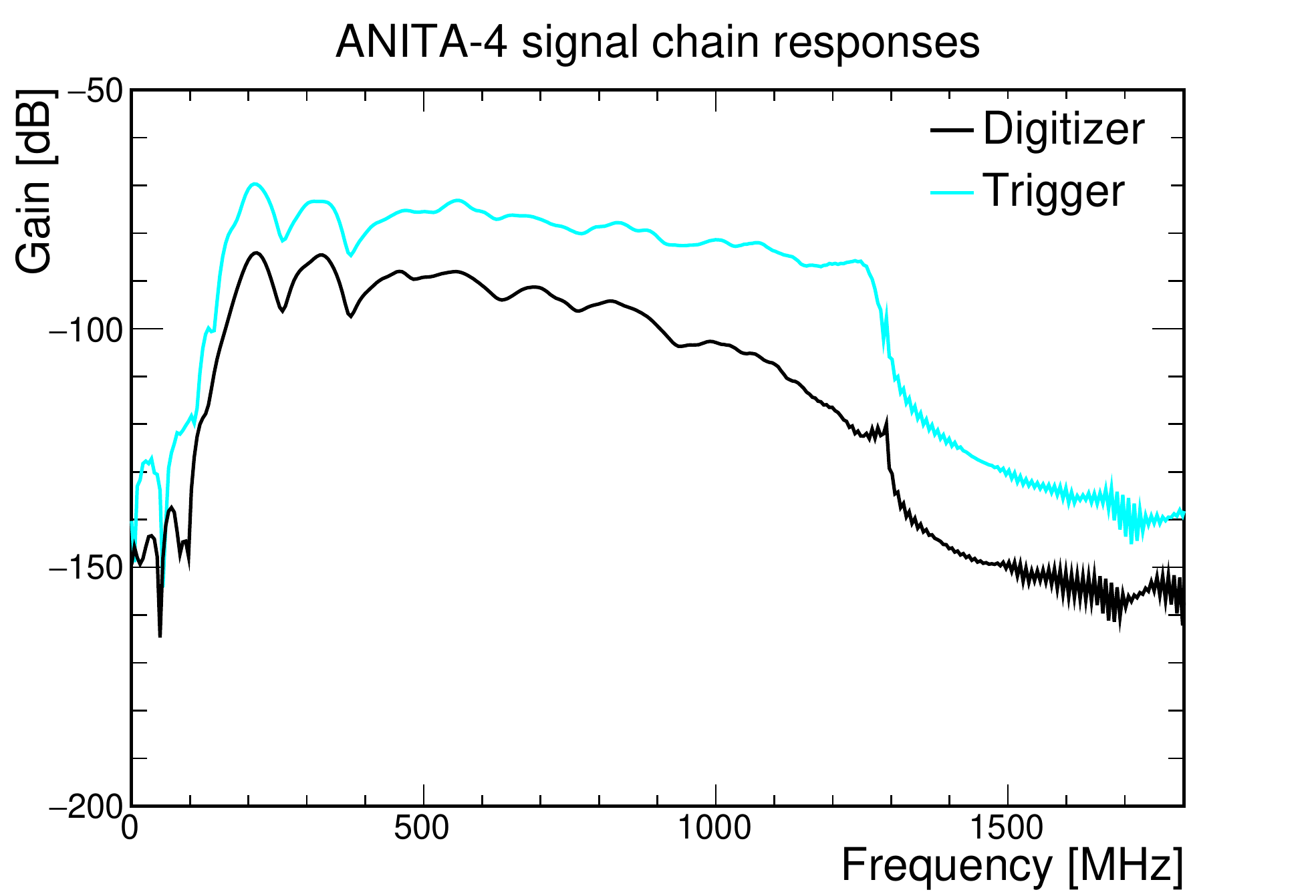}
  \caption{Power spectrum of the trigger and digitizer impulse
    response for a sample channel, for the ANITA-III (left) and ANITA-IV (right) instruments. 
    }
  \label{fig:ANITA_ImpulseResponses}
\end{figure}

The digitizer and trigger responses were measured before the flights
by injecting an RF signal directly into the amplifiers behind the
ANITA antennas and measuring the output of the digitizer and trigger paths; this ensures that the responses include AMPAs and iRFCMs/TUFFs too. 
Figure~\ref{fig:ANITA_ImpulseResponses}~(left) shows the
power spectra of the trigger and digitizer impulse responses for a
sample channel of the ANITA-III payload.
Figure~\ref{fig:ANITA_ImpulseResponses}~(right) shows the
power spectra of the trigger and digitizer impulse responses for a
sample channel of the ANITA-IV payload with the most common filter configuration used during flight.
Thermal noise is generated based on flight measurements (see Section~\ref{subsec:ANITA_thermalNoise}) and added to the waveforms.
Finally, the modeled tunnel diode response (see
Section~\ref{subsec:ANITA_trigger}) is convolved with the signal
waveforms from all the channels, and if an event passes the trigger
logic, the RF signal and truth information about the neutrino
interaction are saved in the final output.

\subsection{Tunable Universal Filter Front-end boards}
\label{subsec:tuffs}
\begin{figure}
  \centering
 \includegraphics[width=0.4\linewidth] {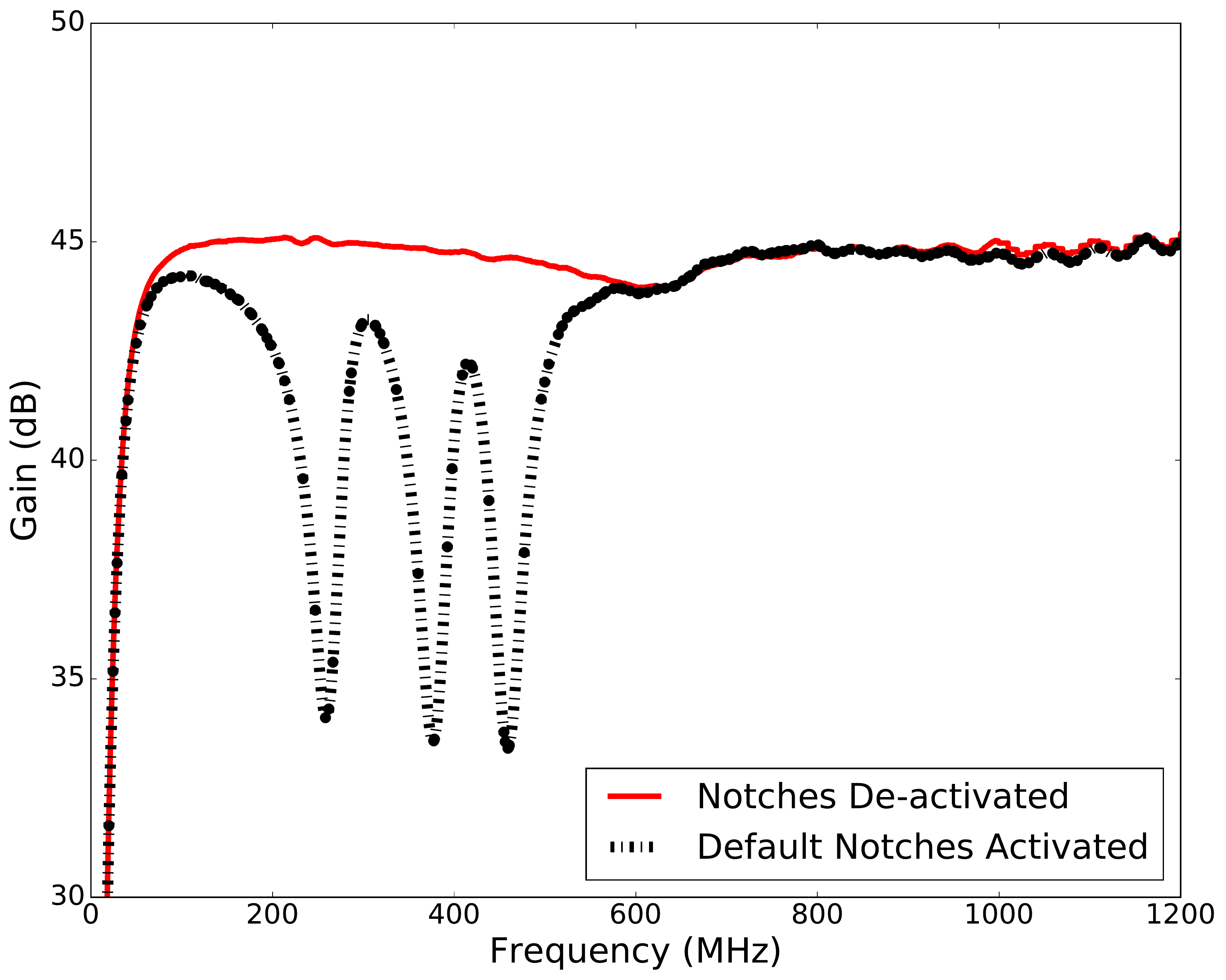} 
 \includegraphics[width=0.45\linewidth] {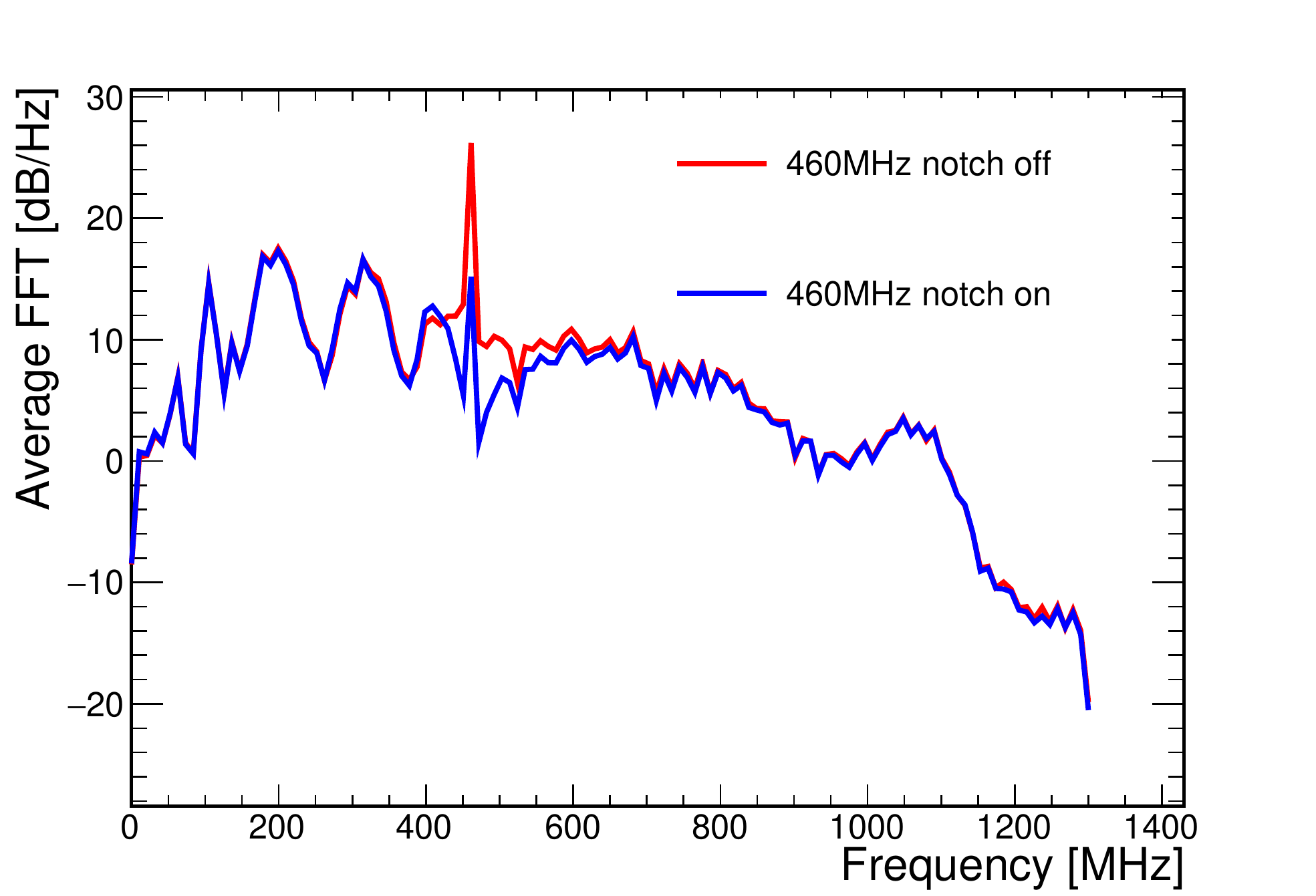} 
  \caption{(Left) Measured TUFF response in frequency domain (dB vs frequency) obtained before the flight. The dips in the response clearly mark the 260, 375, and 460 MHz notches turned on in this configuration and the 13\,dB attenuation at those frequencies is clear~\cite{Allison:2017vtk}.
  (Right) Example of CW noise injected in \icemc with the third notch activated and not activated.}
\label{fig:TUFFs}
\end{figure}

To alleviate the anthropogenic noise observed in ANITA-III that caused
significant amounts of deadtime, ANITA-IV added the Tunable Universal
Filter Front-end, or TUFF, boards~\cite{Allison:2017vtk}.
This board uses up to three notches to attenuate
the gain by a maximum of 13\,dB around each notch
frequency:
the notches are tunable, but default to 260, 375, and 460\,MHz, corresponding to known satellite communications frequencies. 
Whether each notch is activated and at which frequency is called
a "configuration".
There were 16 TUFF boards with six channels each for the 96 total ANITA channels; there were seven unique configurations for the ANITA-IV flight  which are simulated in 
\icemc.
The measured TUFF response for the configuration when all notches are on and at default frequencies is plotted in Figure~\ref{fig:TUFFs}~(left). 
An example of the effect of the third notch switched on or off when 460\,MHz CW noise is simulated is shown in Figure~\ref{fig:TUFFs}~(right).
The measured TUFF response for a given configuration is loaded
into \icemc and convolved with the trigger and digitizer impulse
responses for each channel. 

\subsection{Thermal noise}
\label{subsec:ANITA_thermalNoise}
Modeling thermal noise correctly is crucial to the simulation of ANITA's
sensitivity to neutrino interactions as it affects both the trigger and analysis efficiencies.
An accurate model of the thermal noise makes it possible to simulate both the
accidental noise triggers and the effect of noise fluctuations on the
reconstructed correlation maps used to calculate the event direction in the ANITA analyses~\cite{romero2015interferometric}.

To provide a realistic model of the thermal noise during the ANITA flights, events coming from minimum bias triggers during
a quiet time (a portion of the flight when the payload is not close to any areas of high radio activity, such as active bases or ground pulsers) of the ANITA-III flight are used to produce power spectra in bins of 10\,MHz for each digitizer channel.
As the ANITA-III flight suffered more than the previous flights 
from CW noise coming
from satellites and human bases, frequencies in the ranges
234-286\,MHz and 344-410\,MHz are filtered out in software analysis
using two simple notch filters. 
Data from antennas facing the sun were also excluded.

\begin{figure}[!h]\centering
  \includegraphics[width=.45\linewidth]{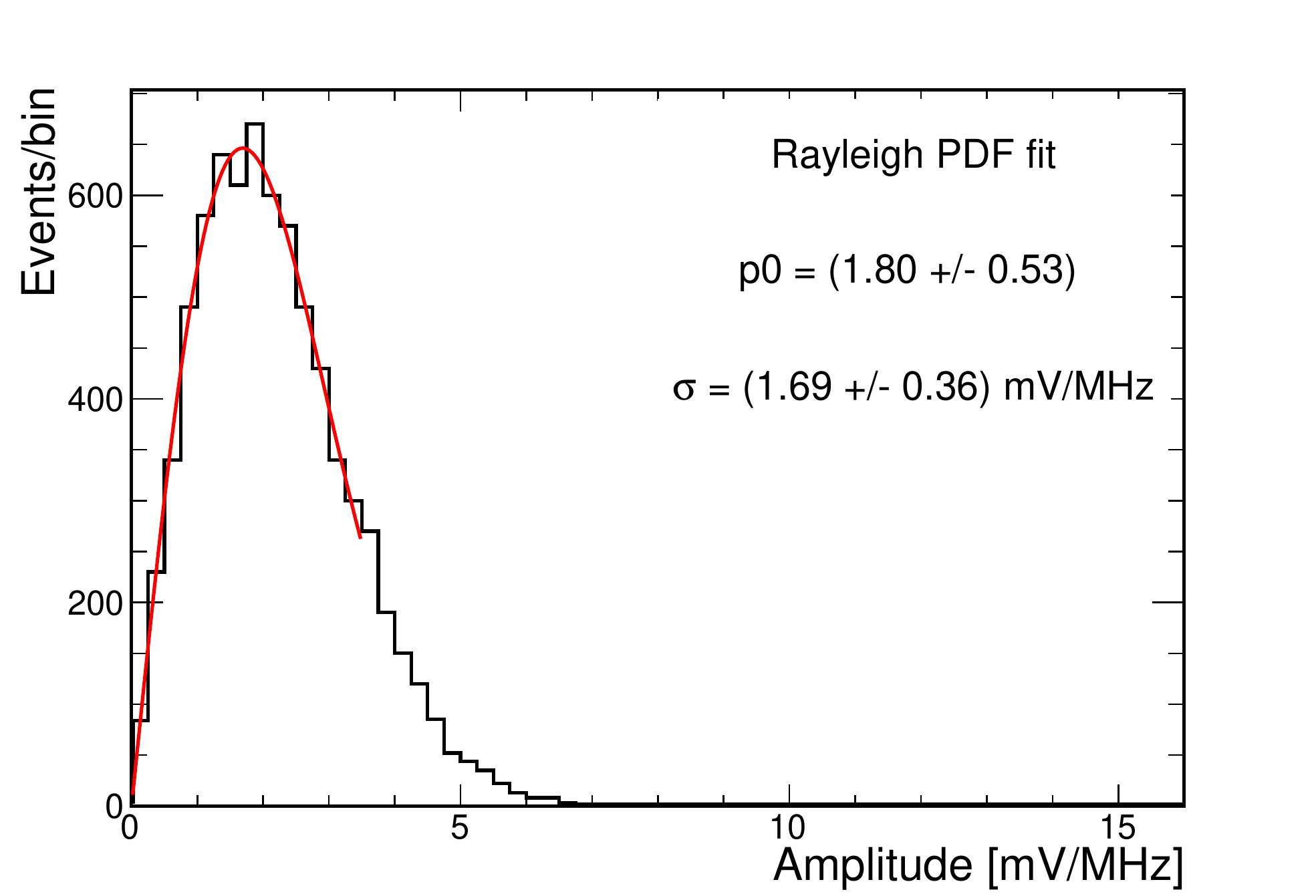}
  \includegraphics[width=.45\linewidth]{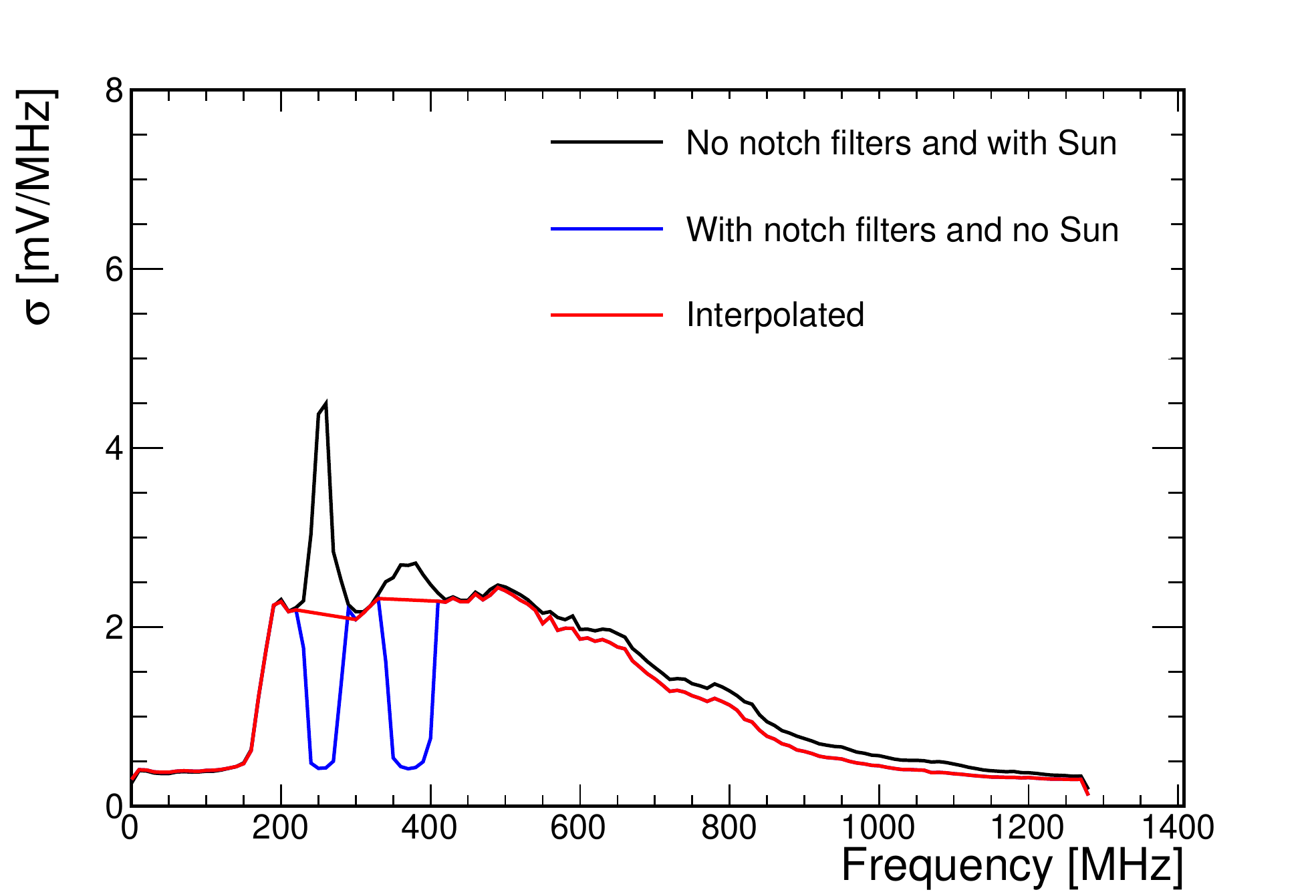}
  \caption{Example of a Rayleigh fit for a sample channel in the
    710.94\,MHz frequency bin (left). The fit is performed only to the rising edge and peak of the distributions, as most CW noise is in the tail. 
Right: fitted amplitude, $\sigma(f_i)$, as a
    function of frequency (interpolating in the frequency range where power is filtered). }
  \label{fig:rayleighFits}
\end{figure}

For each channel and each frequency bin a Rayleigh PDF is fit to the data:
\begin{equation} 
  f(A, \sigma)=\dfrac{A}{\sigma^2}e^{-A^2/(2\sigma^2)} \;,
  \label{eq:rayleigh}
\end{equation}
\noindent where $A$ is the amplitude in the frequency domain (in mV/MHz) and $\sigma$ is the
Rayleigh amplitude in that frequency bin.
As most of the CW noise is in the tail of these
distributions, the Rayleigh fits are performed only to the rising edge and peak of the distributions.
Figure~\ref{fig:rayleighFits} (left) shows an example of Rayleigh distribution fits for a sample channel in the frequency bin centred at 710.94\,MHz.
Events in this sample come from minimum bias triggers during the ANITA-III quiet time.
 
Graphs of the fitted amplitude, $\sigma(f_i)$, as a function of
frequency were produced for each channel (interpolating in the frequency range where power is filtered), as seen in Figure~\ref{fig:rayleighFits} (right).
In \icemc these graphs are used to generate random noise in the
frequency domain for each channel: for each frequency bin $f_i$ the real and imaginary part are randomly extracted from a Gaussian distribution with zero mean and amplitude
$\sigma(f_i)$, the fitted Rayleigh amplitude in that bin.

This noise is added to the signal in the digitizer path.
For the trigger path the noise is re-normalized using the bin-by-bin ratio of the
trigger to digitizer path impulse response in the frequency domain
before adding it to the signal.

Thermal noise for the ANITA-IV simulation is derived from the ANITA-III measurements, accounting for the different electronics response (including the use of low noise amplifiers and variable filter configurations during the flight) for the two missions.
Samples containing only thermal noise are also produced, and they are used in
the main ANITA analyses to test the robustness of our analysis selection.
Subsection~\ref{subsec:validation_flight} details the thermal noise validation for both ANITA flights.

\subsection{Trigger simulation}
\label{subsec:ANITA_trigger}
\begin{figure}[!h]\centering
  \includegraphics[width=.45\linewidth]{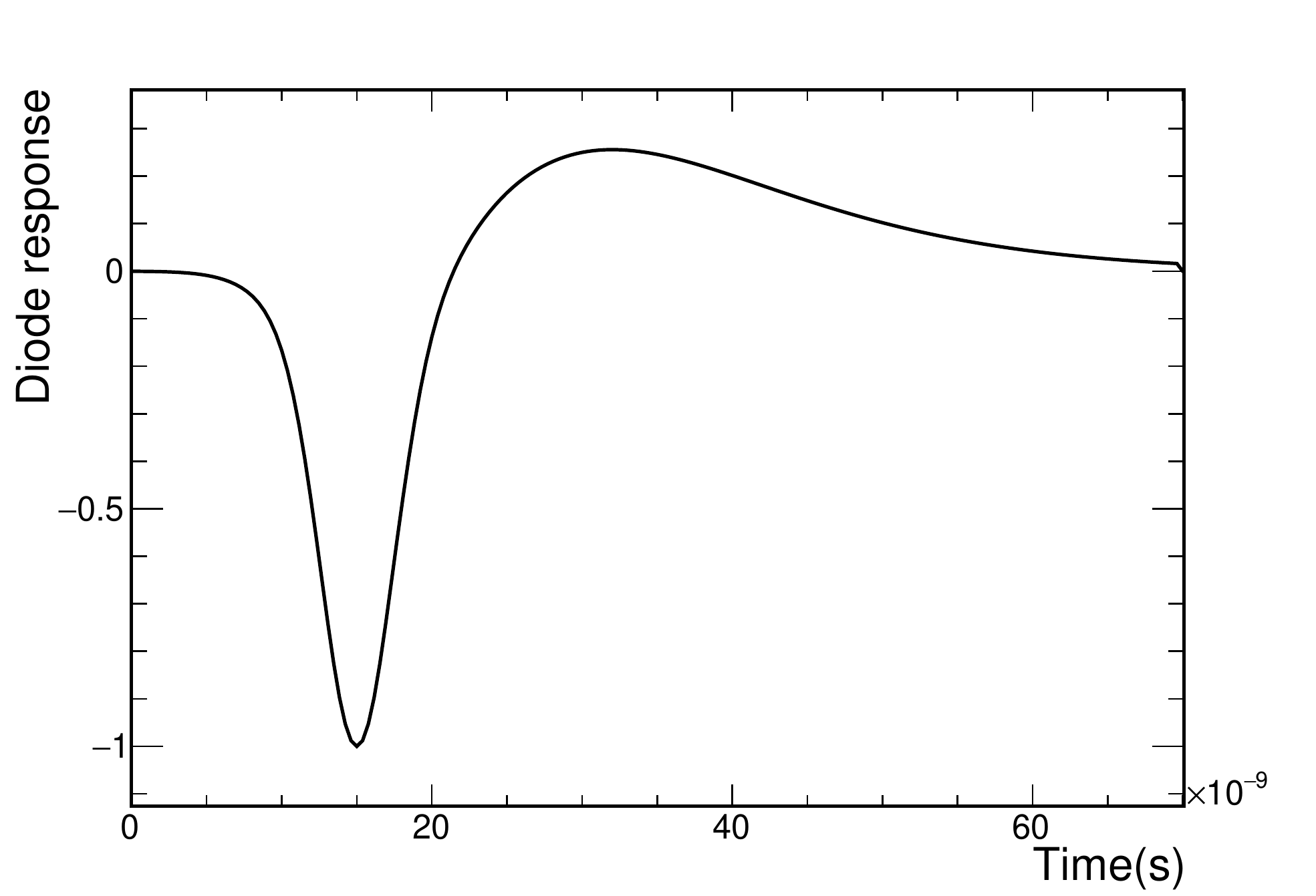}
\caption{Tunnel diode response model used in \icemc.}
  \label{fig:ANITA_diodeModel}
\end{figure}

The simulation models the L0 trigger by passing the
trigger-path signal through a time-domain tunnel diode model and comparing it to the
appropriate threshold from the flight for that channel at the event
time.
The tunnel diode response can be thought of as an integral of the power over about 10\,ns, but the true response is non-trivial.
The shape of the diode response is described by the sum of two negative Gaussians and a positive function that is the product of a quadratic and an exponential:
\begin{equation}
      f(t) = A_1 \cdot e^{-(t-t_0^1)^2/2\sigma_1^2} + 
      A_2 \cdot e^{-(t-t_0^2)^2/2\sigma_2^2} +
      A_3 \cdot \left( t-t_0^3 \right)^2 \cdot
      e^{-(t-t_0^3)/\sigma_3} \;,
    \end{equation}
\noindent where the values for the parameters used for all channels are shown in
Table~\ref{tab:diodeModelParameters};
$f(t)$ is dimensionless.
Figure~\ref{fig:ANITA_diodeModel} shows the diode response model function.

\begin{table}[h!]
\caption{Tunnel diode model parameters used for the full band trigger in ANITA-III and ANITA-IV. The tunnel diode responses for each channel are sufficiently similar that a single set of parameters gives an adequate treatment. }
  \begin{center}
    \begin{tabular}{c|c|c|c} 
      & Value 1 & Value 2 & Value 3 \\
     \hline
      $A$           & -0.8  & -0.2   &  0.00964   \\
      $\sigma$ [ns] &  2.3  &  4.0   &  7.0       \\ 
      $t_0$  [ns]   & 15.0  & 15.0   & 18.0       \\
    \end{tabular}
  \end{center}
  \label{tab:diodeModelParameters}
\end{table}

For each signal, the power as a function of time is calculated as:
\begin{equation}
  P (t) = \dfrac{V(t) \cdot V(t)}{Z} \;,
\end{equation}
\noindent where $V(t)$ is the voltage at time $t$, and $Z=50\,\Omega$ is the system impedance.
The next step is to convolve the waveform power with the diode
response to find the diode output $D(t)$:
\begin{equation}
  D(t) = (f * P)(t) \;.
\end{equation}
For each time bin, the diode output is compared to the channel
threshold, $V_T$, multiplied by the RMS voltage of the diode output coming from pure thermal noise, $V_{RMS}$, following:
\begin{equation} 
  D(t) < V_T \cdot V_{RMS} \;.
\end{equation}
If the diode output is more negative than the threshold multiplied by $V_{RMS}$ at any point, then that channel passes the L0 trigger.

\begin{figure}[!h]\centering
  \includegraphics[width=.45\linewidth]{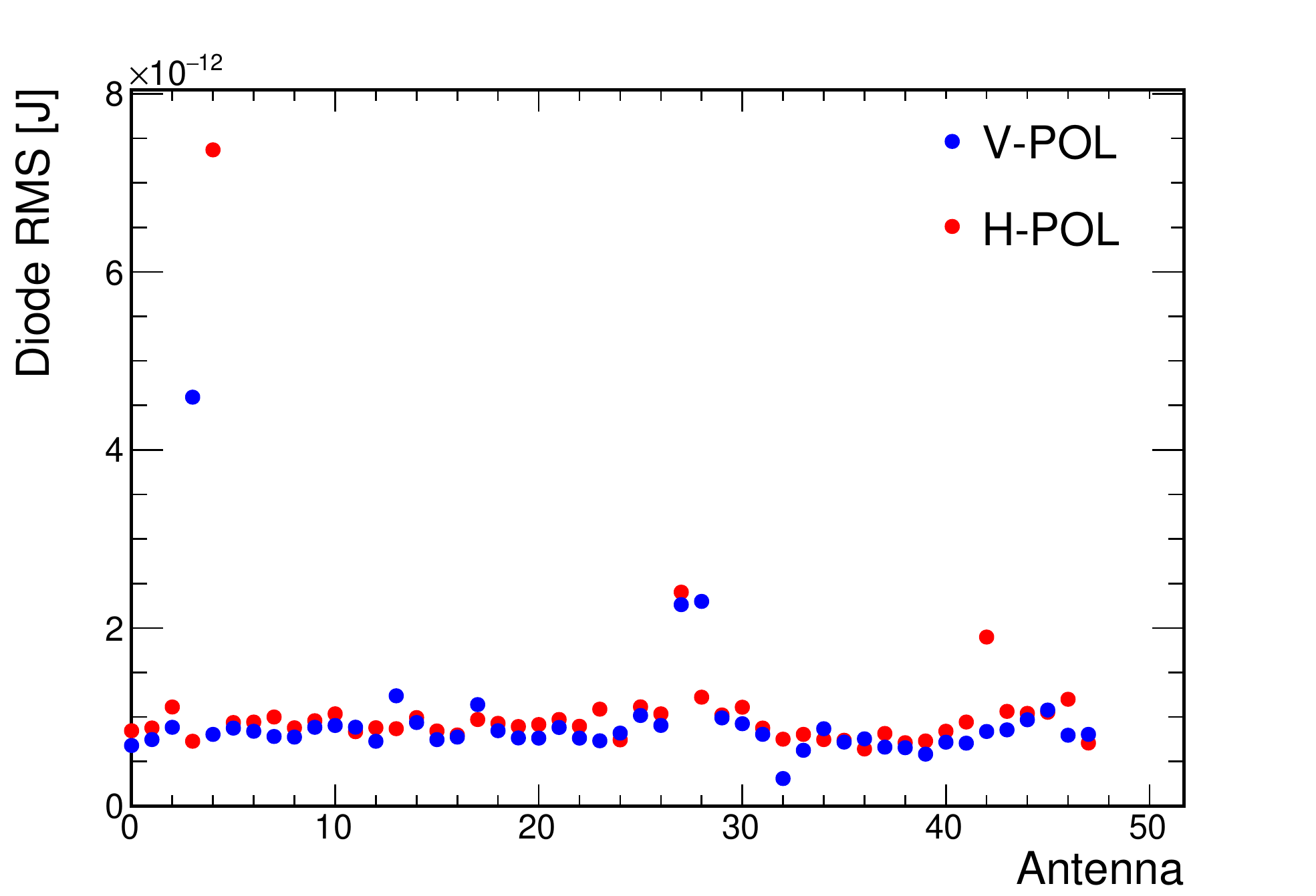}
  \caption{Tunnel diode $V_{RMS}$ for each channel. The two channels with much higher $V_{RMS}$ are the broken channel, and the channel which had an additional filter applied to accommodate the use of an in-flight calibration system.}
  \label{fig:ANITA_diodeRMS}
\end{figure}

At the beginning of a run, $V_{RMS}$ is calculated for each channel 
by simulating 1000 noise waveforms
(Subsection~\ref{subsec:ANITA_thermalNoise}) 
and sampling the diode output at the center of each waveform
(see Figure~\ref{fig:ANITA_diodeRMS}).
In ANITA-III, two channels had very large $V_{RMS}$ associated with them that prevents them from triggering in the simulation: one was broken during the flight, and the other one had an additional filter applied during flight to
allow an in-flight calibration pulser to be used.

The trigger logic for the ANITA-III and ANITA-IV triggers is very similar.
The main difference is that ANITA-IV used 90 degree hybrids to transform H-POL and V-POL signals into LCP and RCP components of waveforms, so the ANITA-IV L1 trigger requires a LCP and RCP L0 coincidence in the same antenna within one 4\,ns clock cycle.
For the ANITA-III instrument an L0 trigger implied an L1 trigger too.
The L2 trigger is formed by a coincidence of two out of three
antennas (top, middle, bottom) in the same azimuthal
sector. As ANITA is intended to look for plane-waves from below, a simple
causal requirement is enforced on the coincidence windows.  An L1 on the bottom
antenna opens a coincidence window open for four FPGA clock cycles (nominally 16\,ns).
Middle and top L1's open the window for three and one-clock cycles (nominally 12 and 4
ns), respectively.
Finally, the global trigger is formed by the coincidence of L2 triggers in
two adjacent azimuthal sectors 
within 3 clock cycles.
For ANITA-III either polarization or both may produce a global trigger. 
Either the L2 or global-trigger may be masked for an
azimuthal sector if the L2 or global rate is too high in the sector.
The actual time-dependent masking status from the flight is used in the simulation.


\section{Validation}
\label{sec:validation}
Different parts of the simulation are validated from measurements
taken before (Subsection~\ref{subsec:validation_lab}) and during
the flights (Subsection~\ref{subsec:validation_flight}).

\subsection{Comparisons with measurements before flights}
\label{subsec:validation_lab}
Before each of the ANITA flights, a series of calibration measurements was
taken at the NASA Long Duration Balloon Facility near McMurdo Station, Antarctica.
These measurements are used to cross-check different parts of the simulation.

\subsubsection{Trigger efficiency scans}
\label{subsec:validation_scans}
\begin{figure}[!h]\centering
  \includegraphics[width=.8\linewidth]{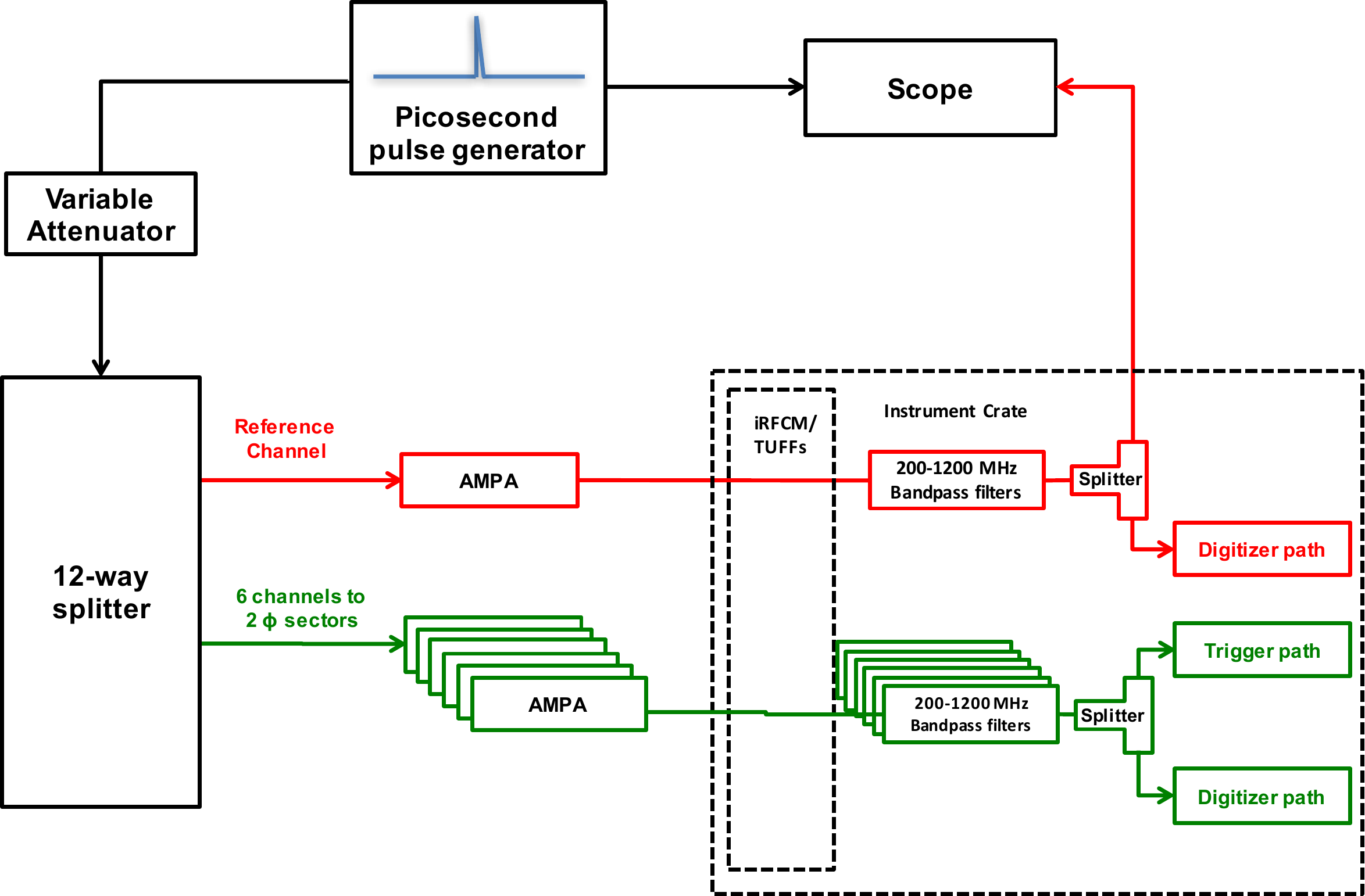}
  \caption{Trigger efficiency scans setup in Antarctica before the ANITA-III flight. A pulser signal is sent to 6 channels (representing 2 azimuthal sectors in one polarization) and is used to measure the trigger efficiency. One reference channel (red line) is sent to the oscilloscope to measure the SNR values at the trigger.}
  \label{fig:scan_setup}
\end{figure}

Trigger efficiency scans are used to measure the ANITA trigger efficiency
for signals with different signal-to-noise ratios (SNRs).
The setup used before the ANITA-III flights is shown in Figure~\ref{fig:scan_setup}. 
A Picosecond Pulse generator is used to produce an RF signal with height 1\,V and FWHM 0.3\,ns. 
One copy of the signal is recorded with an oscilloscope and the other
is passed through attenuators, and a 12-way splitter.
Six of these channels (corresponding to
two azimuthal sectors in one polarization) are inserted directly into the amplifiers behind the ANITA antennas, and follow the standard trigger and digitizer paths.
One reference channel is inserted into the amplifier and routed from the trigger path to the oscilloscope to measure the SNR values at the trigger.

To simulate the trigger efficiency scan setup, the signals
measured at the oscilloscope (see Figure~\ref{fig:scan_setup}) 
are inserted (with appropriate attenuation) into the same six channels as used
in the hardware efficiency scans.
These go through the trigger/digitizer path and produce ANITA
data-like outputs.

\begin{figure}[!h]\centering
  \includegraphics[width=.45\linewidth]{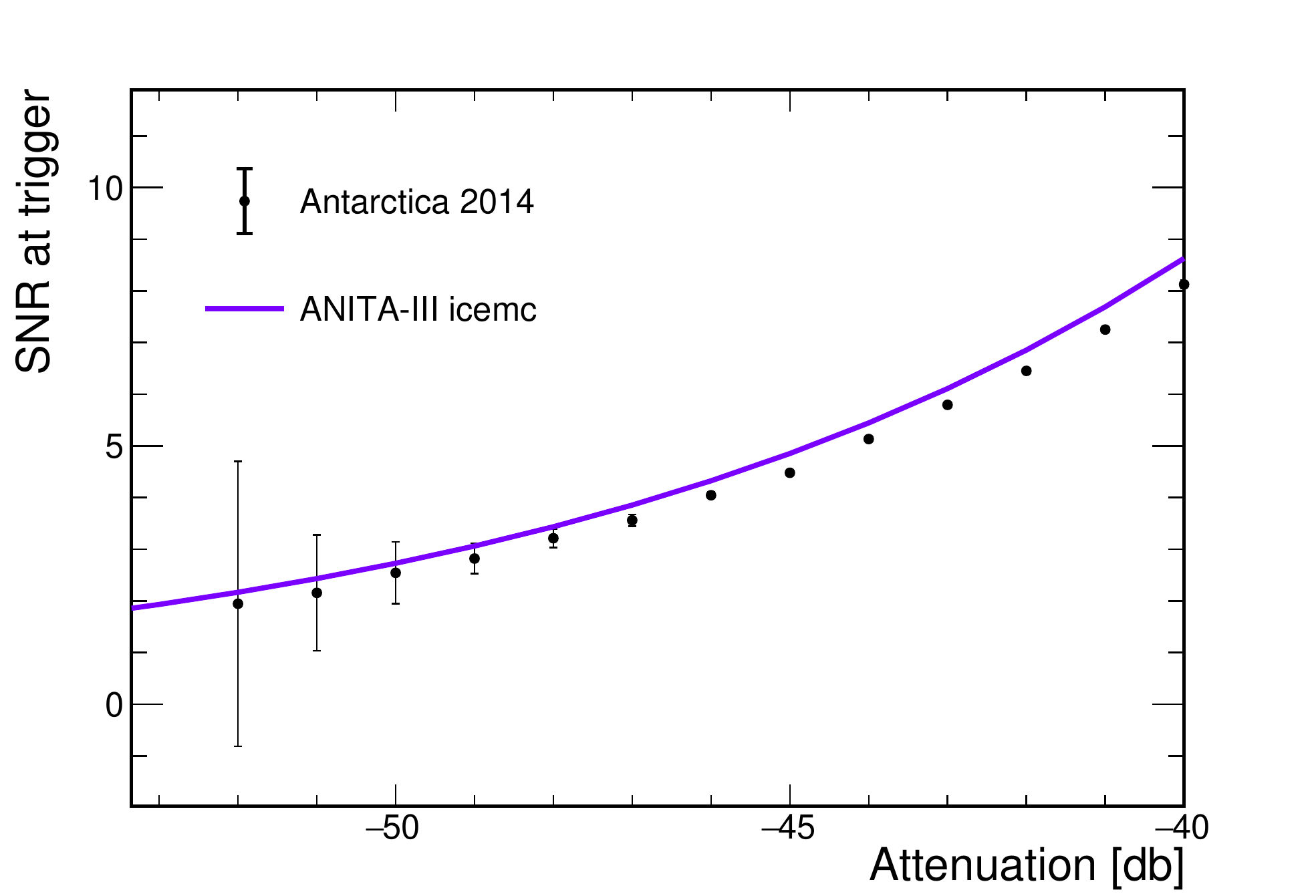} 
  \includegraphics[width=.45\linewidth]{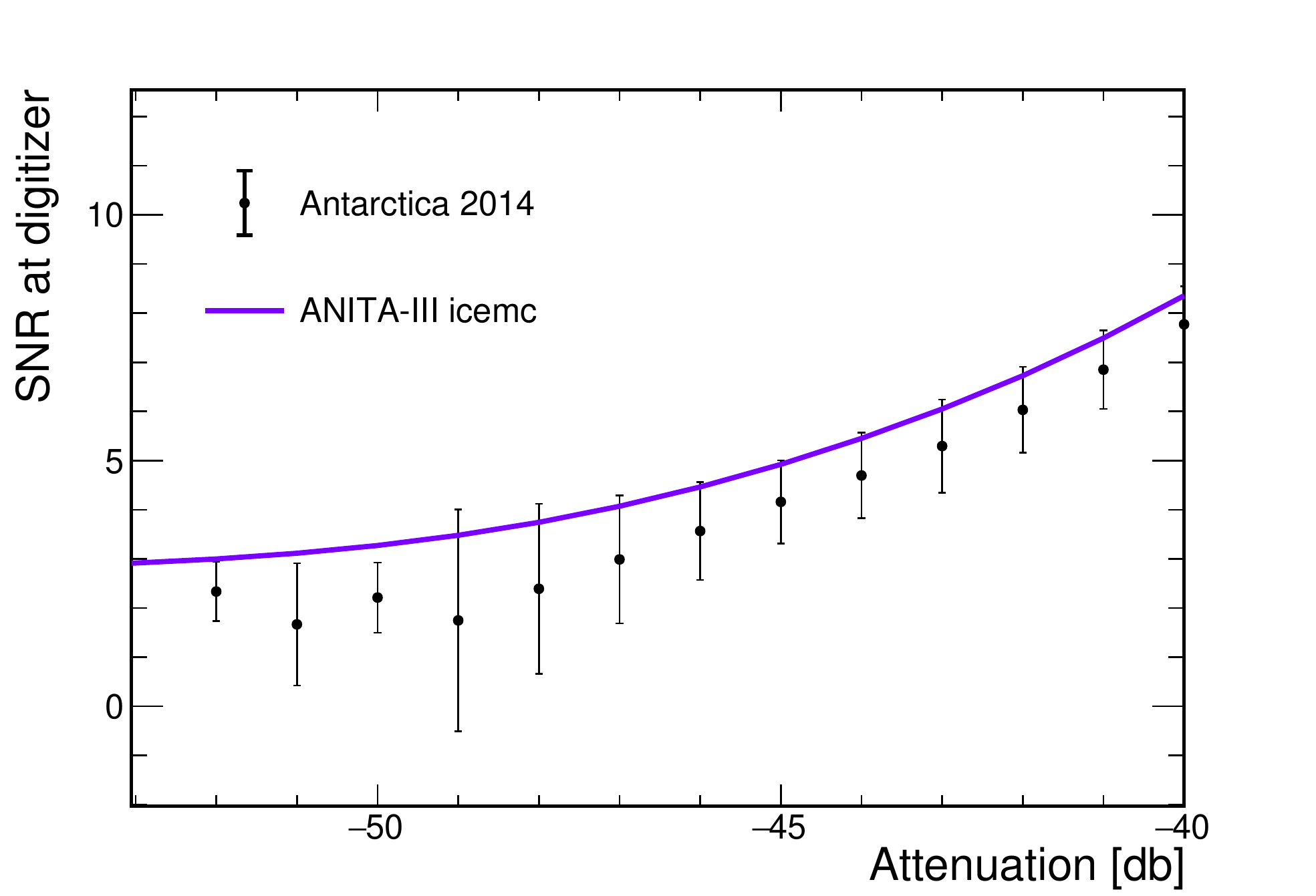}
  \caption{SNR measured at the trigger (left) and digitizer (right) as
  a function of the variable attenuation applied during the trigger
  efficiency scans. The data points were measured in 2014 in Antarctica prior to the
  ANITA-III flight.
}
  \label{fig:scan_snr}
\end{figure}

The signals are recorded at the trigger and the digitizer paths, and
are used to validate the simulation.
Figure~\ref{fig:scan_snr} shows a comparison of the measured SNR in
data and simulation at the trigger (left) and at
the digitizer (right) as a function of the variable attenuation used
during the scan.
The SNR is calculated as the ratio of half of the peak-to-peak of the
time domain signal to the RMS of the last part of the waveform.
The measured SNR in the digitizer path generally has larger uncertainties compared to the SNR in the trigger path, because the ANITA digitizer has on average a
2.6\,GHz sampling rate, whereas the trigger path is measured with a
fast oscilloscope. 

\begin{figure}[!h]\centering
  \includegraphics[width=.45\linewidth]{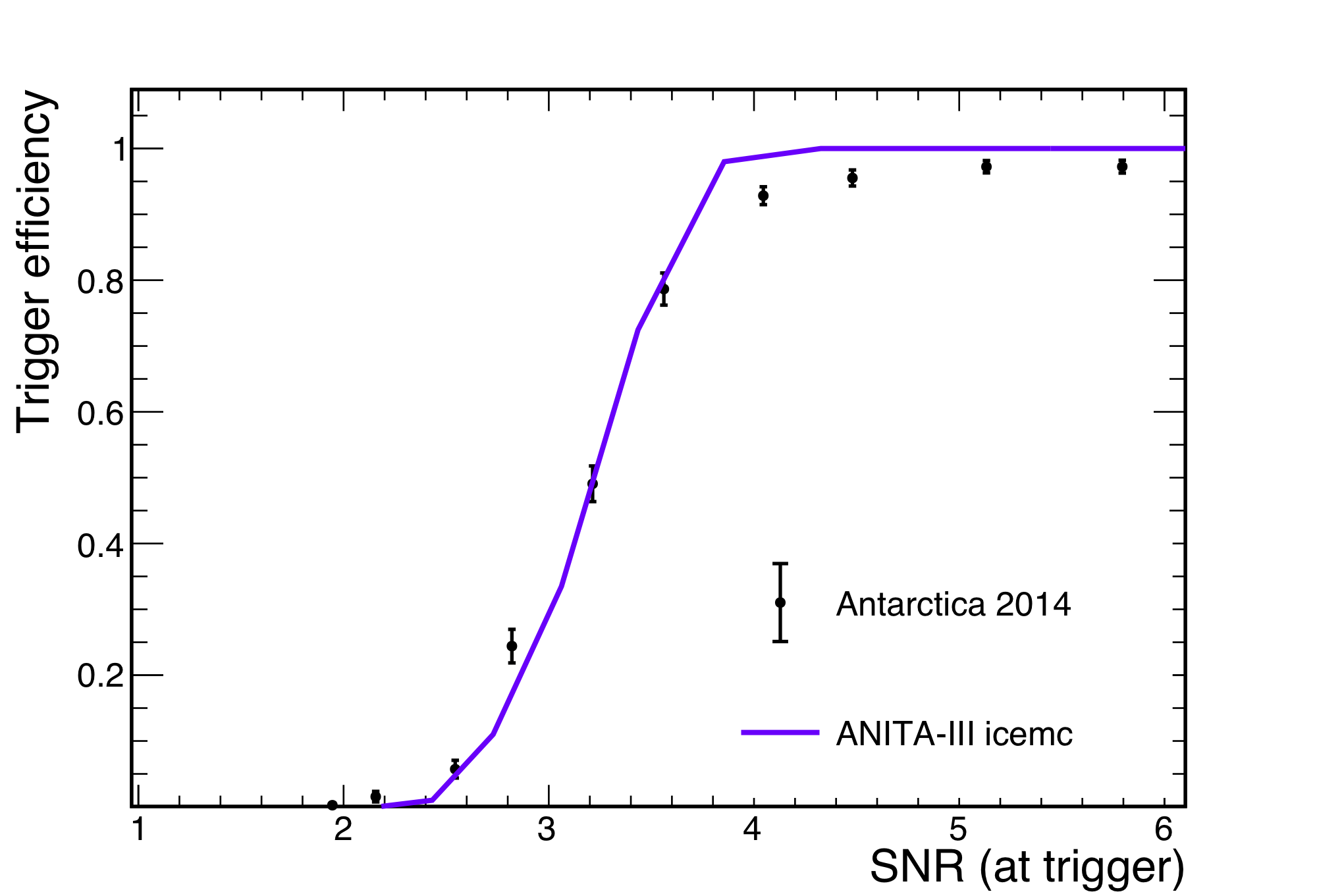}
    \includegraphics[width=.45\linewidth]{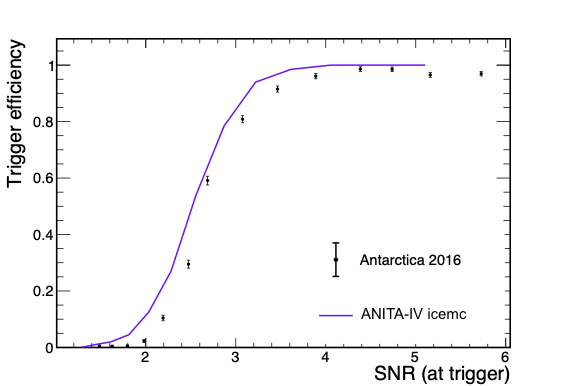}
  \caption{Comparison between data and simulation of a trigger efficiency scan for the ANITA-III (left) and ANITA-IV (right) instruments. 
    The trigger efficiency is plotted as a function of the SNR
    measured at the oscilloscope for the data, and the SNR estimated in the trigger path for the simulation. 
    }
  \label{fig:scans}
\end{figure}

Figure~\ref{fig:scans}~(left) shows a comparison between data and simulation of a trigger efficiency scan for the ANITA-III payload.
The trigger efficiency is plotted as a function of the SNR
measured in the trigger path.

Before the ANITA-IV flight, a similar set of measurements was collected.
A data and simulation comparison of the trigger efficiency for the ANITA-IV instrument is shown in Figure~\ref{fig:scans}~(right).

\subsection{Comparisons with flight measurements}
\label{subsec:validation_flight}
Data taken during the ANITA-III and ANITA-IV flights are used to validate 
the thermal noise, the trigger efficiency to a ground pulser, and the pointing reconstruction.

\subsubsection{Thermal noise validation}
\label{subsec:ANITA_validation_thermalNoise}
\begin{figure}[!h]\centering
  \includegraphics[width=.45\linewidth]{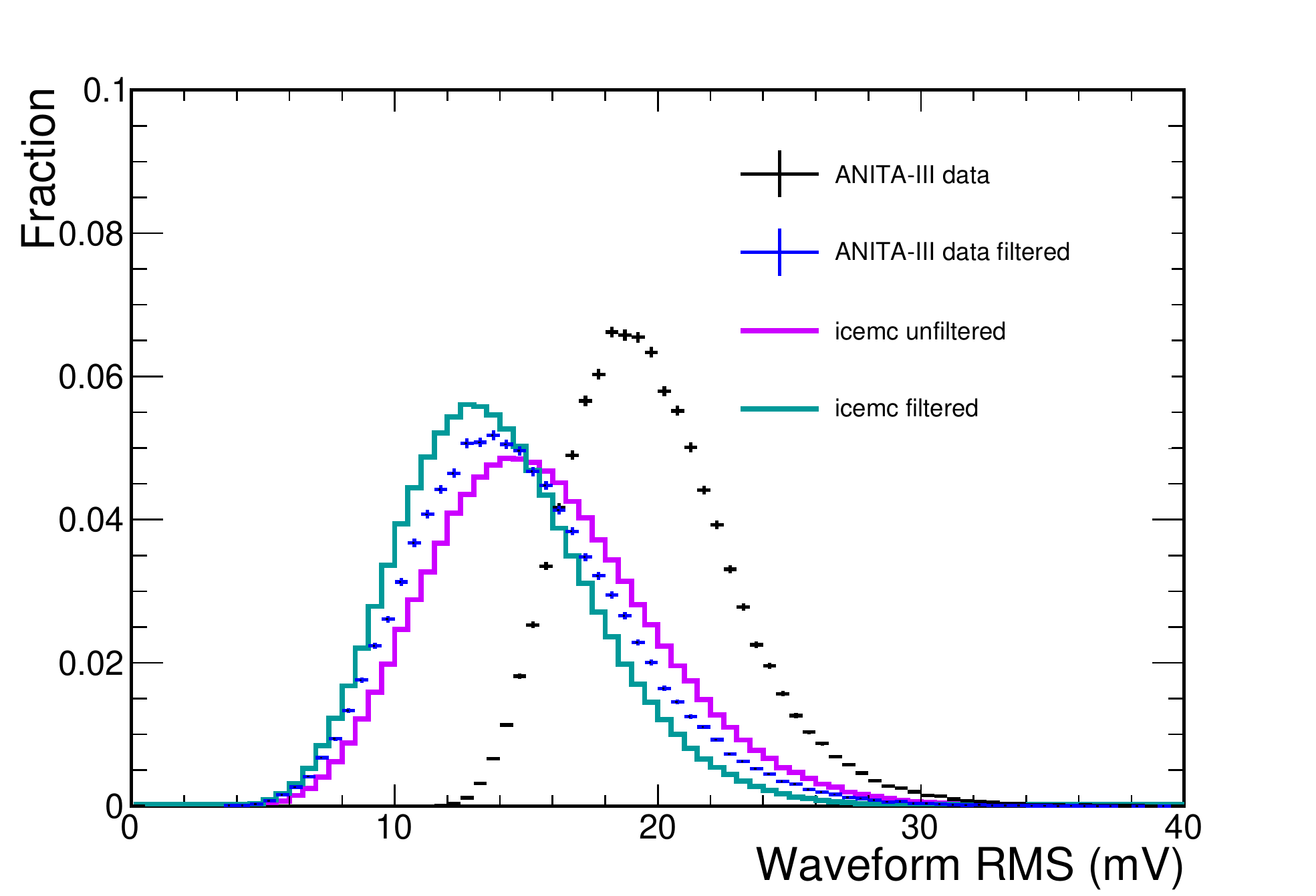}
  \includegraphics[width=.45\linewidth]{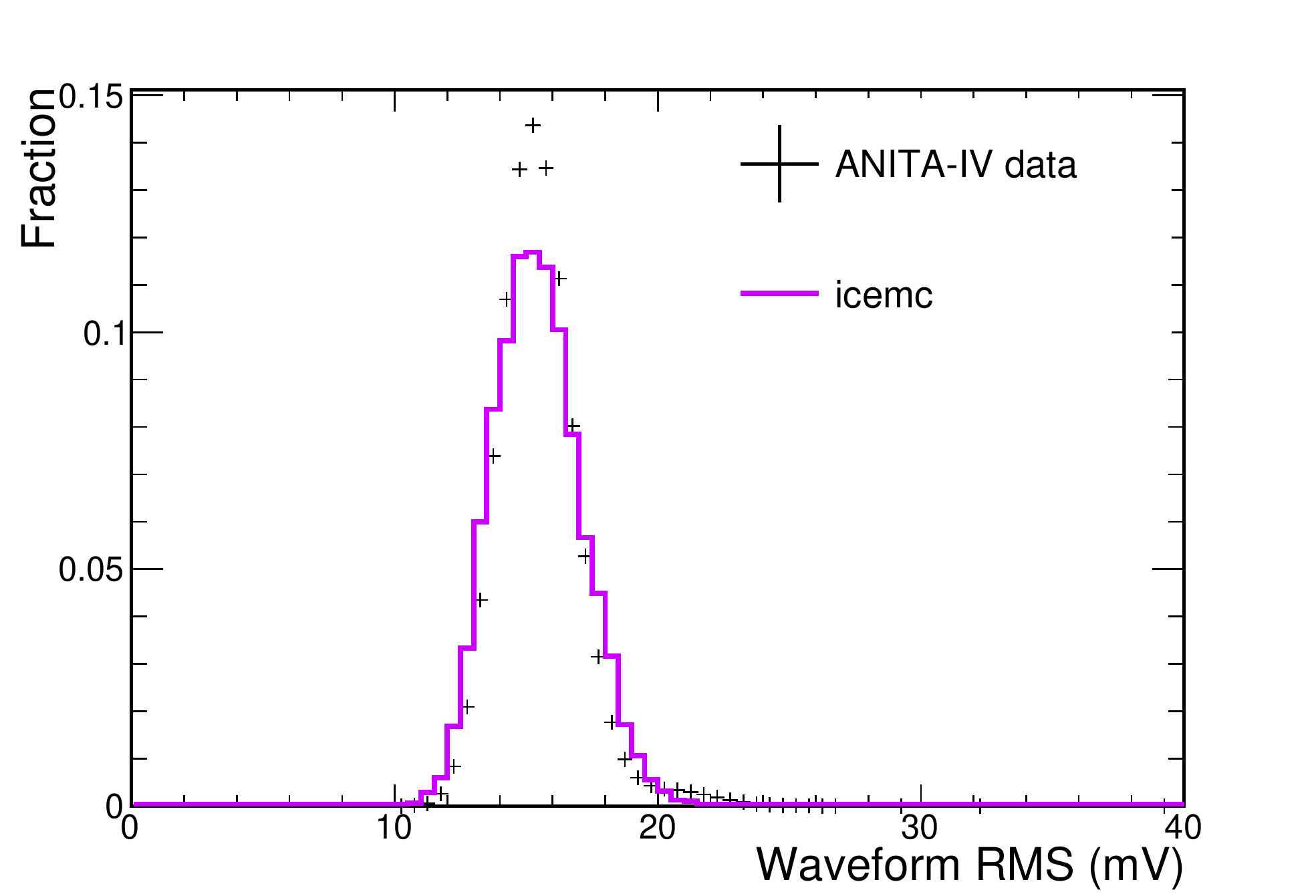}

\caption{Thermal noise validation. Comparison of RMS of waveforms for an \icemc simulation and a relatively quiet period of the ANITA-III (left) and ANITA-IV (right) flights.
    The ANITA-III flight was affected by strong CW noise that for a better comparison has been filtered out.
    These distributions are normalized to equal area to better compare their shape.
  }
  \label{fig:RMSwaveform}
\end{figure}

The simulation of thermal noise is validated using distributions of
the RMS of the simulated waveforms compared to a relatively quiet time
during the flights.
Figure~\ref{fig:RMSwaveform} shows the voltage RMS of noise-only waveforms produced
in \icemc compared with the ones coming from a quiet time during
the ANITA-III (left) and ANITA-IV (right) flights.

As the ANITA-III data contained a significant amount of CW
contamination, a fair comparison of the data and simulation thermal noise is done after applying two notch filters to both datasets, improving the data and simulation agreement.
The remaining differences are due to CW noise in the data that
could not be simply removed with two notch filters.

The ANITA-IV payload was less affected by CW noise, as the TUFF boards were directly filtering out noisy frequencies, and the requirement of the coincidence between LCP and RCP signals to form a trigger ensured that only linearly polarized signals triggered the payload.
A direct comparison between the measured and simulated noise is seen in Figure~\ref{fig:RMSwaveform}~(right), where the better data-simulation agreement for ANITA-IV, than for ANITA-III, is due to the reduced CW noise in the ANITA-IV datasets.
The simulation slightly overestimates the ANITA-IV thermal noise, but these small differences do not impact the experiment sensitivity.

As a second consistency check, the ANITA-III simulated thermal
noise is also used to produce Rayleigh fits (as described in
Subsection~\ref{subsec:ANITA_thermalNoise}) and shows complete overlap with the fits to the
ANITA-III quiet time.

\subsubsection{WAIS pulser model}
\label{subsec:wais}

As an additional validation, we model the calibration pulser that was located at the West Antarctic Ice Sheet (WAIS) field camp during the ANITA-III and ANITA-IV flights. The WAIS pulser consisted of a 6-kV FID GmbH brand~\cite{fid} pulser that generates a broadband impulse and drives a horizontally polarized antenna. The pulser was triggered on the GPS second with a known delay, permitting a measurement of the  ANITA trigger efficiency while the payload is in view of the pulser.

The antenna used at WAIS is a custom design based on a quad-slot model, for which the slots of the antenna are parallel to the ground. The antenna was installed $\sim$1~m below the surface of the snow. Similar to a discone, the bottom portion of the antenna acts as a reflector, and the angles of the taper on both the reflector and the upper cone tune the orientation of the peak gain. This design is a scaled down version of the VHF antenna used as a low frequency extension to ANITA-III. 

\begin{figure}
\centering
\includegraphics[width=\linewidth]{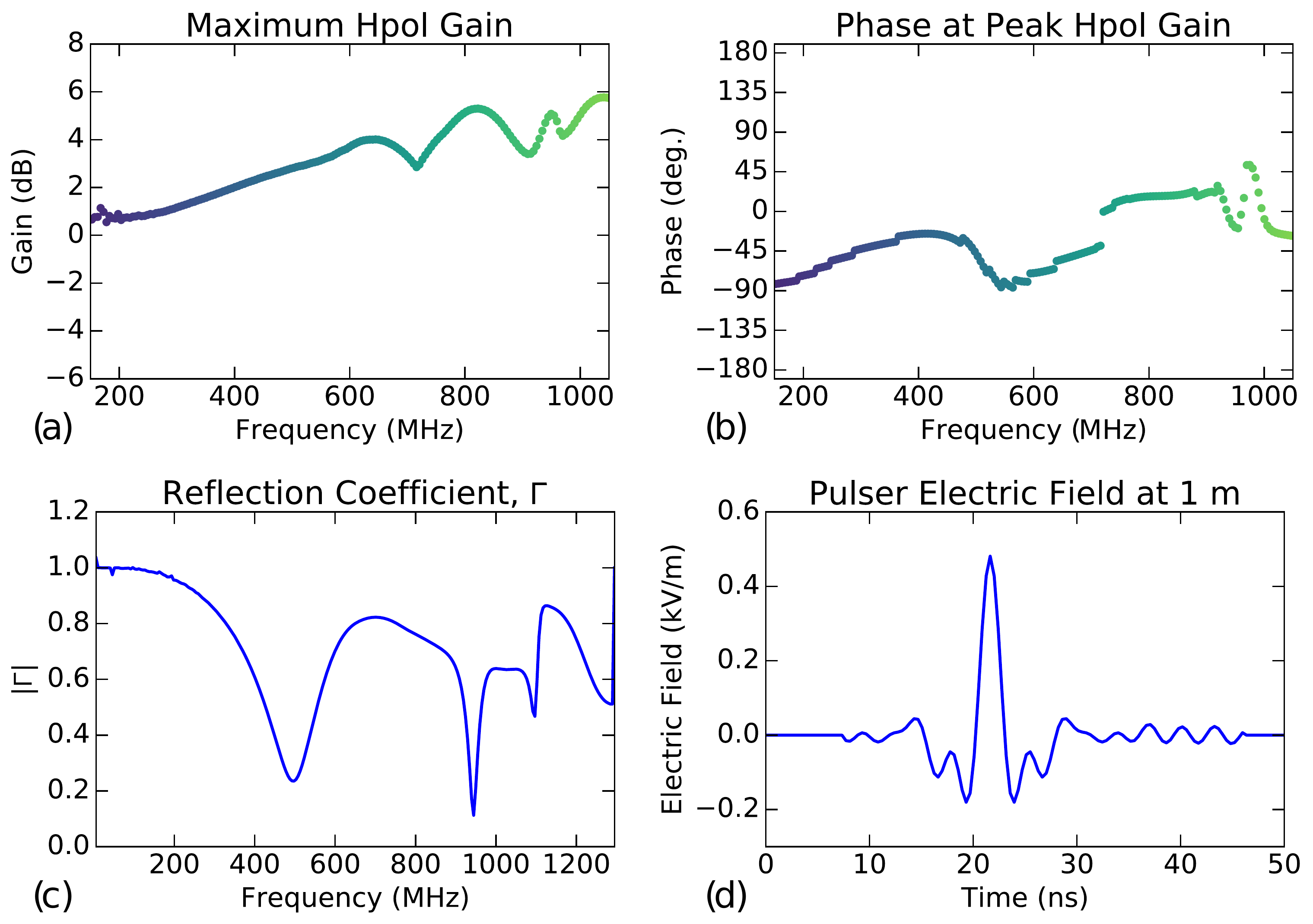}
\caption{Characterization of the ANITA-III horizontally polarized antenna used in the WAIS pulser.}
\label{fig:waisPulser}
\end{figure}

The WAIS antenna response was modelled using NEC antenna modeling software~\cite{nec}. Figure~\ref{fig:waisPulser} shows the peak gain for each frequency, the associated phase at that frequency, and the reflection coefficient, $\Gamma(f)$, from this antenna model. We then model the electric field (Figure~\ref{fig:waisPulser}d) generated by the WAIS pulser by convolving the NEC model of the antenna with measurements of the voltage generated by the FID pulser on an oscilloscope. The electric field at 1\,m from the pulser, $E(f)$, results from relating the power density radiated by the antenna to the power density generated by the pulser with a characteristic impedance $Z_{c}$ and voltage $V(f)$ at the pulser, accounting for loss from imperfect antenna matching between the antenna and the pulser, the magnitude and phase of the gain, $G(f)$, propagation loss, and the impedance of free space, $Z_0$:

\begin{equation}
|E(f)| = \sqrt{\frac{|V(f)|^2}{8 Z_c} (1 -|\Gamma(f)|^2 ) \frac{Z_0}{2\pi (1~\textrm{m})^2} G}
\end{equation}

\begin{figure}
\centering
\includegraphics[width=.5\linewidth]{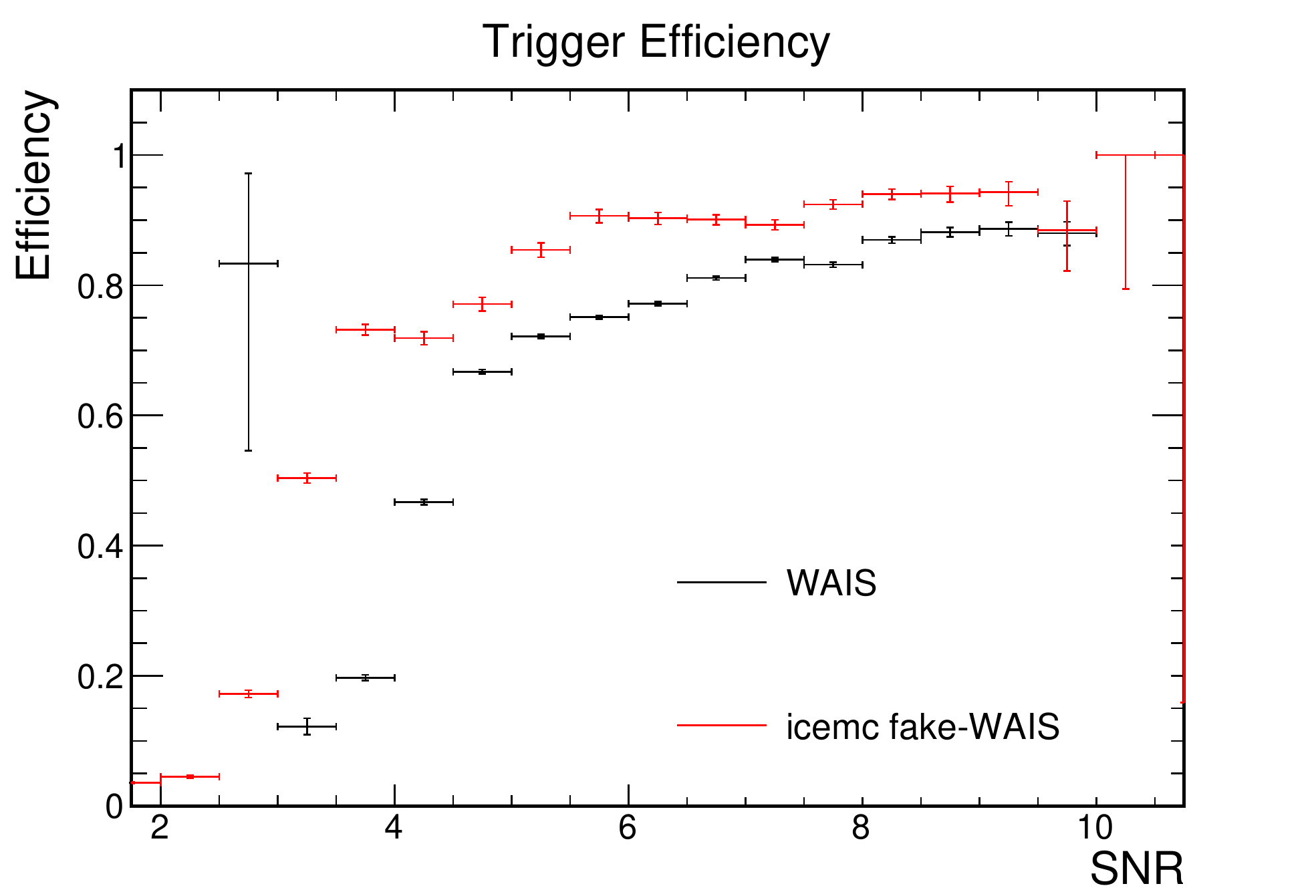}
\caption{ANITA-III trigger efficiency to a pulser coming from the WAIS divide station as a function of SNR.}
\label{fig:waisEff}
\end{figure}

The simulation treats the electric field shown in Figure~\ref{fig:waisPulser} as originating from a source at the location of the WAIS pulser. Figure~\ref{fig:waisEff} shows the efficiency of the ANITA-III payload to generate a trigger from pulses coming from WAIS divide (the linear border between the ice flows at WAIS). The simulation efficiency does not asymptote to 1 at high SNR, as it also includes inefficiencies due to the channel masking and the instrument dead time. 
The data and simulation efficiency are used as systematic uncertainties in the calculation of the sensitivity of the experiment.

\subsubsection{Reconstruction validation}
\label{subsec:ANITA_validation_reconstruction}
A large sample of simulated
neutrinos is produced following the cosmogenic neutrino flux arising from a mixed cosmic-ray composition as modeled by Kotera et al.~\cite{kotera} and used to validate the simulated pointing resolution.
Simulated waveforms from different triggering channels are
cross-correlated to form a pointing map in the ANITA payload
coordinates, azimuth ($\phi_{\mathrm{meas}}$) and elevation
($\theta_{\mathrm{meas}}$). 
The peak of the correlation map is then compared to the expected
azimuth ($\phi_{\mathrm{theory}}$) and elevation
($\theta_{\mathrm{theory}}$), calculated from the true neutrino interaction point.
The angular resolution in azimuth and elevation of the simulated neutrinos is comparable to the one measured using the WAIS pulser during the flights.
Details of our reconstruction and analysis can be found in our
previous publications~\cite{ANITA1paper,ANITA2paper,romero2015interferometric}.

\section{ANITA sensitivity}
\label{sec:results}

The simulation files produced with \icemc are used by the ANITA analysts to tune analysis cuts, check the rate of accidental clustering, and also to simulate sources like Gamma Ray Bursts and Active Galactic Nuclei.
Finally, the simulation is used to calculate the experiment sensitivity.

\subsection{Acceptance}
\label{subsec:acceptance}
The ANITA collaboration uses \icemc to calculate the experiment volumetric acceptance \effvol, following:
\begin{equation}
  \label{eq:volacceptance}
  \effvol = \dfrac{n_{{pass}}(E)V_0\Omega}{N(E)} \;,
\end{equation}
\noindent where
 $n_{{pass}}(E)$ is the weighted (see Section~\ref{sec:weights}) number of events that pass the trigger at a given energy $E$,
 $V_0$ is the volume of ice in Antarctica viewed by ANITA,
$\Omega$ is $4\pi$ steradians, and 
 $N(E)$ is the number of neutrino events thrown by each simulation at that energy $E$.

The ANITA acceptance (\effarea) is calculated following:
\begin{equation}
  \label{eq:acceptance}
  \effarea = \dfrac{\effvol}{\ell_{{int}}(E)} \;,
\end{equation}
\noindent where
 \effvol is the volumetric acceptance and
 $\ell_{{int}}(E)$ is the average interaction length in that energy bin. 
The interaction length is calculated following:
 \begin{equation}
   \label{eq:intlength}
    \ell_{{int}}(E_\nu) =   \dfrac{M_{NUCL}}{\sigma(E) \rho_{H_2O} } \;,
  \end{equation}
  
\noindent where
 $M_{NUCL}$ is the atomic nuclear mass ($1.66\cdot 10^{-27}$\,kg),
$\sigma(E)$ is the neutrino cross section for $\nu$ charged-current interactions, 
and
 $\rho_{H_2O}$ is the density of water (1000\,kg/m$^3$), 
Currently the neutrino cross section is calculated using either the
Reno {\it et al.}~\cite{reno2005high}
or the Connolly {\it et al.}~\cite{PhysRevD.83.113009} parametrizations.
The latter is the default.

\begin{figure}[!h]\centering
  \includegraphics[width=.45\linewidth]{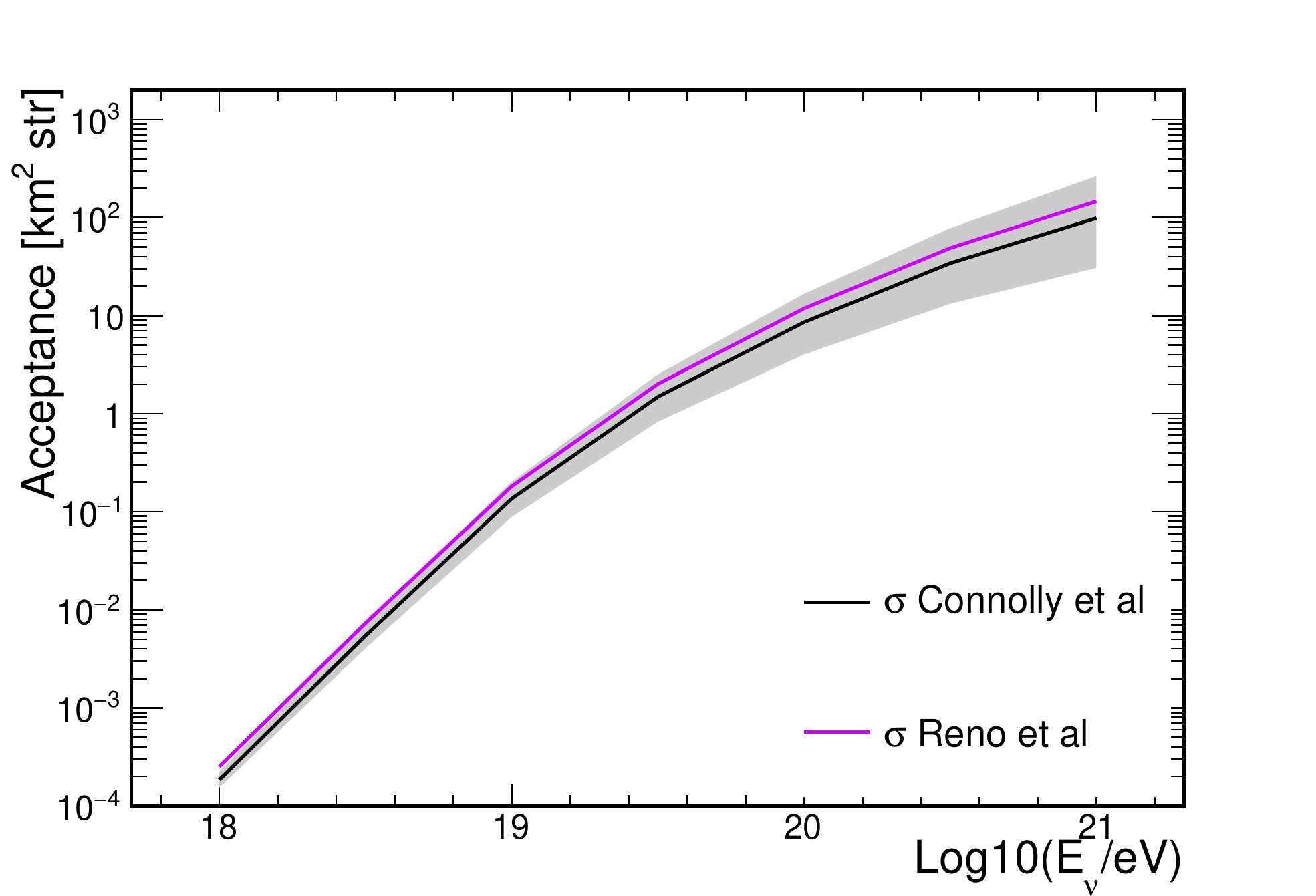}
  \includegraphics[width=.45\linewidth]{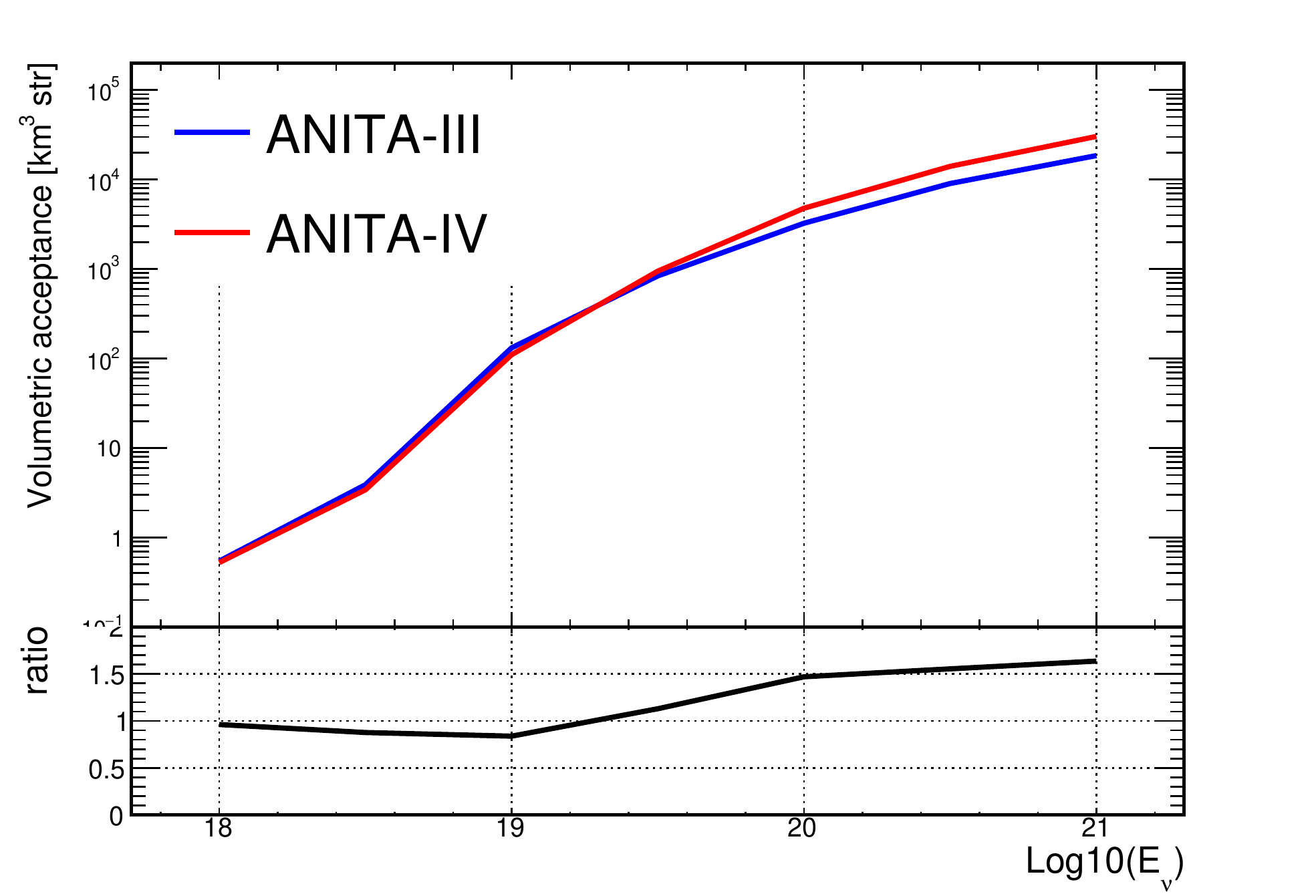}
  \caption{(Left) ANITA-III acceptance for different cross section parametrizations: Connolly et al.~\cite{PhysRevD.83.113009} and Reno et al.~\cite{reno2005high}. (Right) ANITA-III and ANITA-IV volumetric acceptance comparison as a function of energy.}
  \label{fig:acceptanceVSxsec}
\end{figure}

Although recent neutrino cross section measurements by the IceCube
Collaboration reached the multi-TeV
scale~\cite{aartsen2017measurement,bustamante2017measurement}, 
above this energy there are
only cross section measurements with an uncertainty of a factor $\pm 5$ up to
300\,TeV.
The current theoretical models can extrapolate the cross sections up to $10^{21}$\ev~\cite{PhysRevD.83.113009,reno2005high},
but the associated uncertainties are
large, and the impact on the ANITA acceptance is
non-negligible.
Figure~\ref{fig:acceptanceVSxsec}~(left) shows the effect of changing the
cross section parametrization on the ANITA-III acceptance.
The nominal Connolly {\it et al.} parametrization is compared to the upper and lower bound set by
Reference~\cite{PhysRevD.83.113009}, and to the alternate
parametrization suggested by Reference~\cite{reno2005high}.

\begin{figure}[!h]\centering
  \includegraphics[width=.45\linewidth]{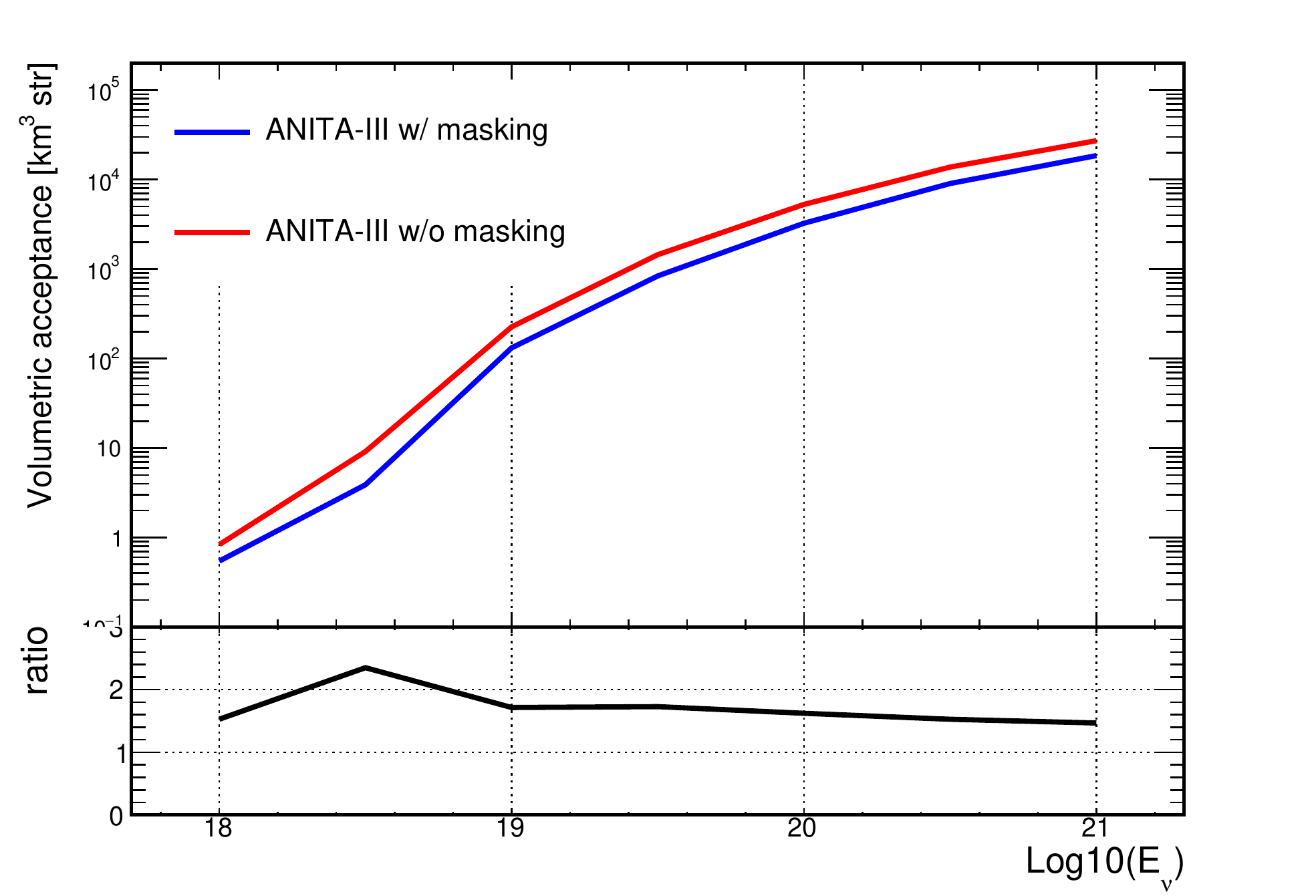}
  \includegraphics[width=.45\linewidth]{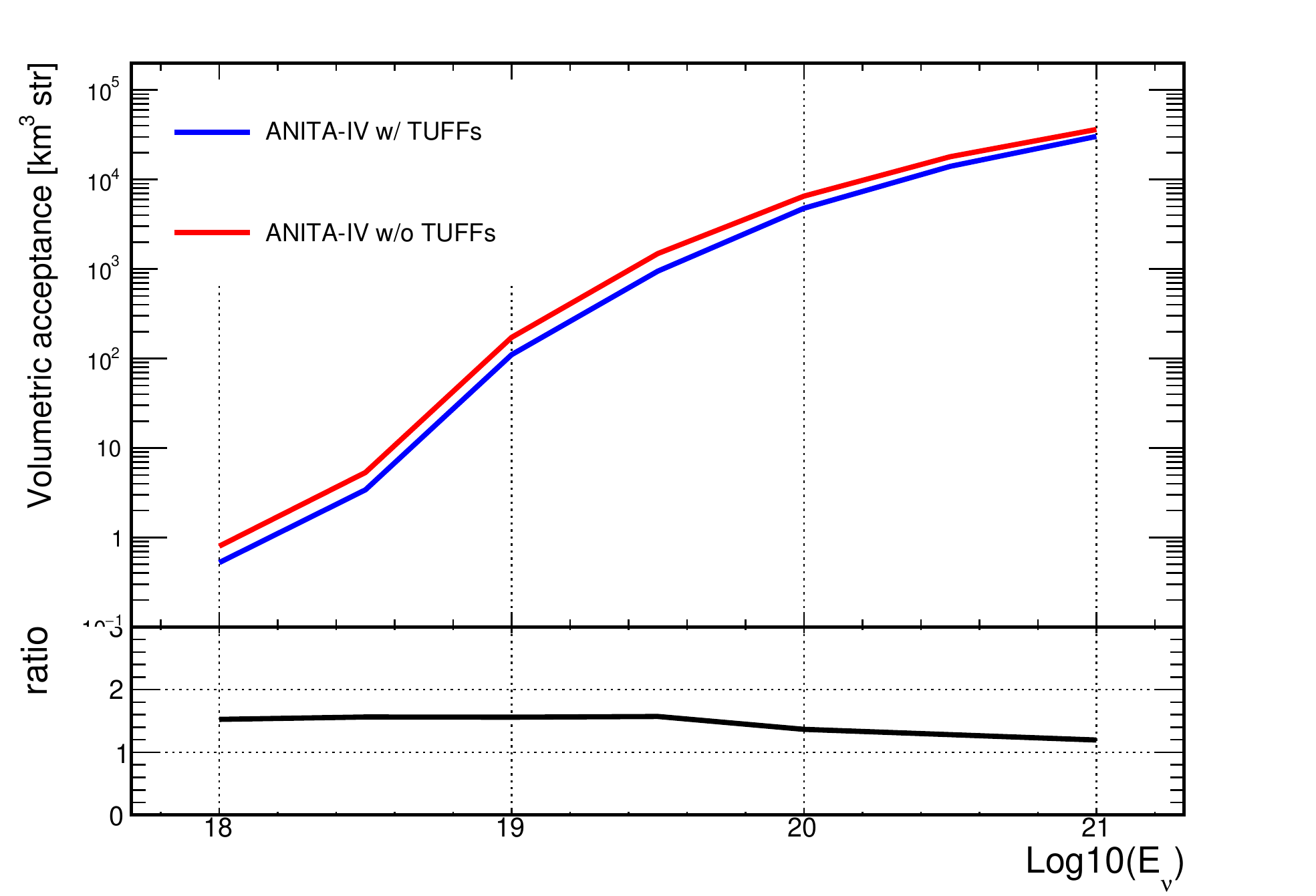}
  \caption{(Left) Comparison of ANITA-III volumetric acceptance with and without channel masking.
  (Right) Comparison of ANITA-IV volumetric acceptance with and without TUFF boards.}
  \label{fig:acceptanceVariations}
\end{figure}

Figure~\ref{fig:acceptanceVSxsec}~(right) shows the ANITA-III and ANITA-IV volumetric acceptances and their ratio.
The ANITA-IV hardware improvements (use of better low noise
amplifiers, the tunable notch filters to avoid CW noise, and
the use of LCP-RCP trigger coincidences to avoid satellite noise) 
enabled ANITA-IV to increase the volumetric acceptance by 50\% at the
highest energy.

Volumetric acceptances are also used to evaluate the impact of different hardware choices.
ANITA-III was highly affected by CW and satellite noise and had to employ channel masking throughout the entire flight to avoid overloading the trigger. Figure~\ref{fig:acceptanceVariations}~(left) shows that channel masking reduced the ANITA-III sensitivity by more than 50\% across all energy bins.
ANITA-IV used the TUFF boards and the coincidence of LCP-RCP triggers to avoid CW and satellite noise, hence only used channel masking for a relatively small fraction of the flight. 
Nonetheless, the TUFF boards filtered out significant fractions of the frequency range and had a non-negligible impact on the sensitivity of ANITA-IV:  Figure~\ref{fig:acceptanceVariations}~(right) shows the impact of the use of the TUFF boards on the ANITA-IV sensitivity, which was reduced by 25--50\%.

\begin{figure}[!h]\centering
  \includegraphics[width=.45\linewidth]{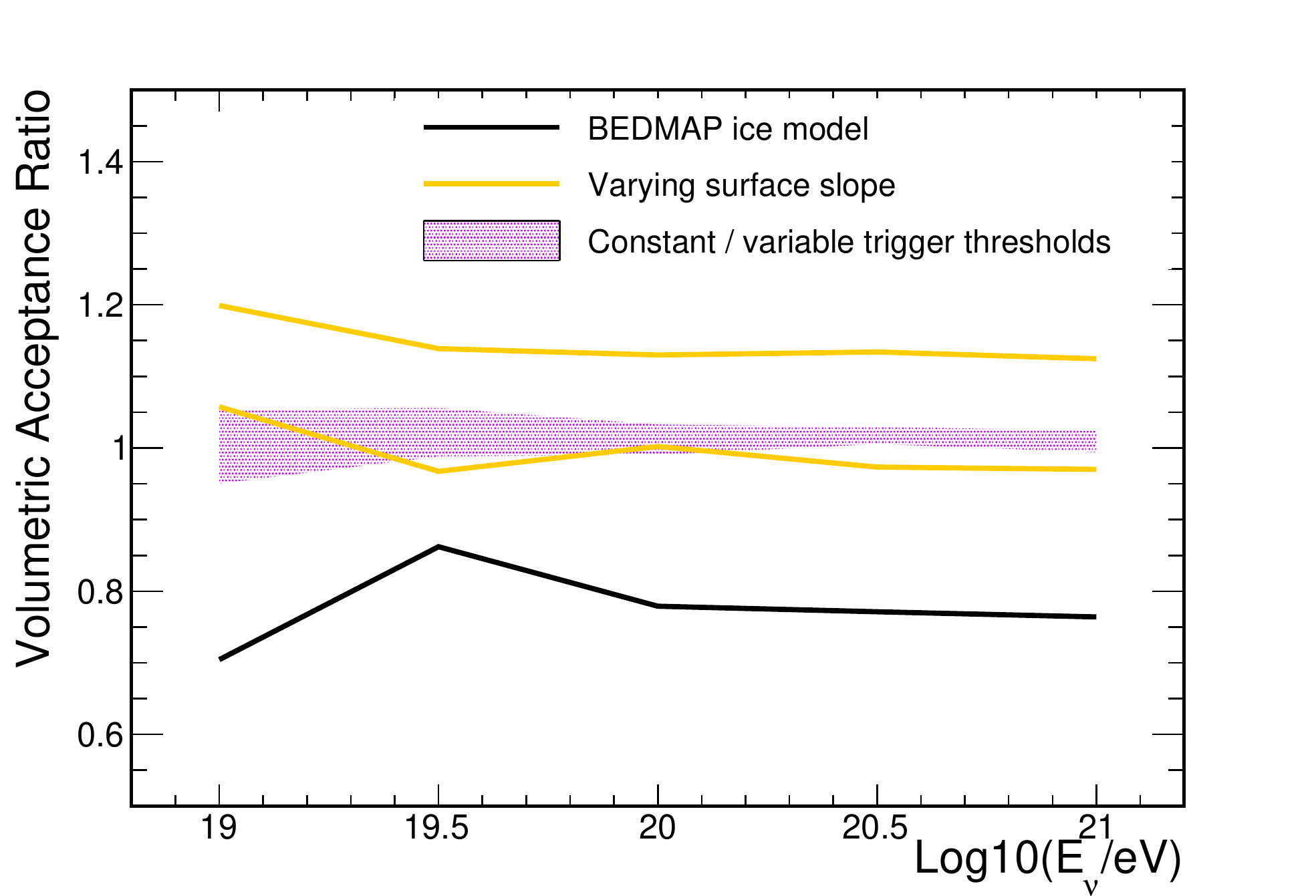}
  \caption{Ratio of ANITA-IV volumetric acceptances as a function of energy found varying different \icemc parameters. The solid black line shows the variation coming from using BEDMAP instead of CRUST 2.0 as Antarctica ice model.
  The orange solid lines are calculated by setting the surface slope inclination at 0 or 2.4\% (double the nominal).
  The upper and lower boundaries of the violet area correspond to the change in volumetric acceptance that would have resulted if the trigger thresholds  had  been  set  respectively  at  their  minimum  or  maximum  values for the entire flight rather than using the dynamic threshold.
   }
  \label{fig:moreAcceptanceVariations}
\end{figure}

Figure~\ref{fig:moreAcceptanceVariations} shows the variation of the ANITA-IV volumetric acceptances coming from different \icemc parameter variations.
The black solid line shows the effect of using BEDMAP instead of CRUST 2.0 as Antarctica ice model; the more finely binned BEDMAP map results in a roughly 20\% lower acceptance over all energies.
Future versions of the simulation will use BEDMAP2~\cite{fretwell2013bedmap2}, with improved ice bed, surface and thickness datasets for Antarctica.
The orange solid lines shows the effect of varying the random surface inclination, the two cases shown are 0 and 2.4\%, where the latter is double the nominal value.
The violet area shows the change in acceptance resulting from using a range of fixed trigger thresholds (from the maximum to  the  minimum  threshold) instead  of  the  dynamic threshold actually used during the flight.

\subsection{Limit}
\label{subsec:limit}
The projected 90\% confidence level on the diffuse neutrino flux is set by using:
\begin{equation}
\left( \dfrac{Ed^4N}{dE dA d\Omega dt}\right)_{lim} =
\dfrac{s_{up}}{ T \cdot \epsilon_{ana} (E_\nu) \cdot \effarea \cdot \Delta} \, ,
\end{equation}
\noindent
where
$s_{up}$ is the upper (one-sided) limit for the mean of a Poisson
variable given 0 observed events in the absence of background for 90\%
CL, 
$T$ is the live time 
(17.4 days for ANITA-III and 24.25 days for ANITA-IV), 
$\epsilon_{ana}(E_\nu)$ is the neutrino analysis efficiency,
\effarea is the experiment acceptance as a function of the neutrino
energy, and $\Delta=4$ is a model-independent factor following Reference~\cite{PhysRevD.73.082002}.

\begin{figure}[!h]\centering
 \includegraphics[width=.65\linewidth]{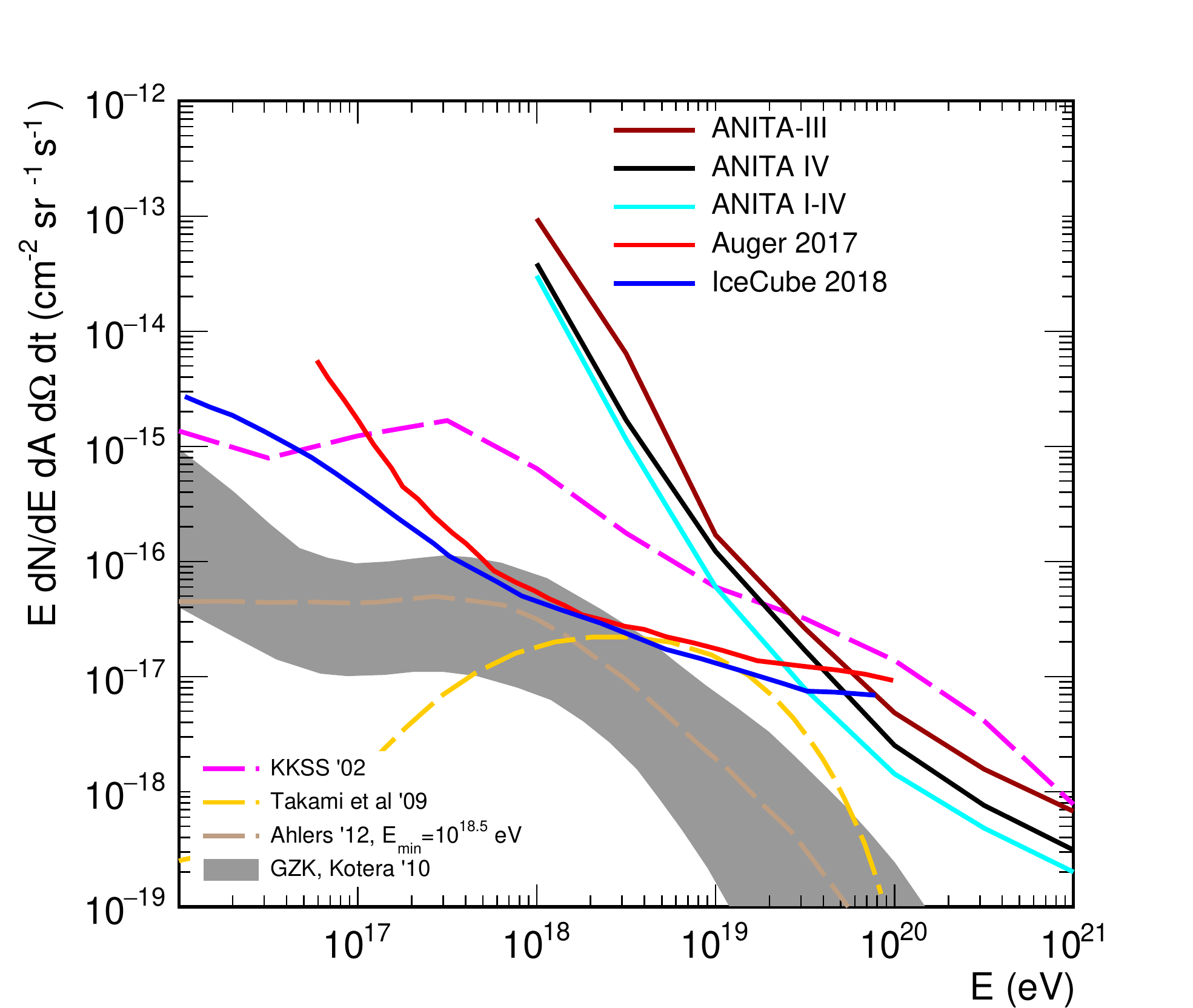}
 \caption{
 ANITA-III and ANITA-IV limit on the all flavor diffuse UHE neutrino flux and a combined limit from ANITA I-IV, as calculated in References~\cite{anita3cosmogenic,anita4cosmogenic}.
 The most recent UHE neutrino limits
from the Auger~\cite{auger2017} and IceCube~\cite{icecube2018} experiments, and
four cosmogenic neutrino models~\cite{kkss2002,takami2009,ahlers2012,kotera2010cosmogenic} are also displayed. 
}
 \label{fig:sensitivity}
\end{figure}

Figure~\ref{fig:sensitivity} shows the ANITA-III, ANITA-IV and ANITA\,I-IV limits as calculated in References~\cite{anita3cosmogenic,anita4cosmogenic}.
These are compared to the latest constraints coming from the IceCube~\cite{icecube2018} and Auger experiments~\cite{auger2017}, as well as
four cosmogenic neutrino models~\cite{kkss2002,takami2009,ahlers2012,kotera2010cosmogenic}.

\section{Summary and Future improvements}
\label{sec:future}
The \icemc Monte Carlo simulation program is used for the
simulation of ultra high energy neutrino interactions in the Antarctic
ice and their detection by the ANITA experiment.
The data taken before or during the ANITA flights are used to validate the
simulation.
The latest ANITA-III, ANITA-IV and ANITA\,I-IV performance is provided in the form of the
acceptance and energy dependent neutrino limit.
 
Future versions of the program will include the possibility to use extensive air showers inputs from ZHAireS~\cite{alvarez2012monte}; improved ice bed, surface and thickness datasets from BEDMAP2~\cite{fretwell2013bedmap2};
refined ice properties modeling and surface roughness effects; the contribution of the sun to the thermal noise; and CW noise as measured during the ANITA-III and ANITA-IV flights.
We are also working towards expanding the framework so that it can be used by multiple radio experiments based in Antarctica.

\section{Acknowledgements}
\label{sec:ackowledgements}
We would like to thank the National Aeronautics and Space Administration and the National Science Foundation, especially the Office of Polar Programs.  We would especially like to thank the staff of the Columbia Scientific Balloon Facility and the logistical support staff enabling us to perform our work in Antarctica.  We are deeply indebted to those who dedicate their
careers to help make our science possible in such remote environments.
This work was supported by the Kavli Institute for Cosmological Physics at the University of Chicago.  Computing resources were provided by the University of Chicago Research Computing Center and the Ohio Supercomputing Center at The Ohio State University.
A. Connolly and S. Wissel would like to thank the National Science Foundation for their support through CAREER awards 1255557 and 1752922, respectively.
A. Connolly would also like to thank the National Science Foundation for their support through the single-PI grant GRT00049285.
O. Banerjee and L. Cremonesi's work was also
supported by collaborative visits funded by the Cosmology and Astroparticle
Student and Postdoc Exchange Network (CASPEN).
The University College London group was also supported by the Leverhulme Trust. The Taiwan team is supported by Taiwan's Ministry of Science and Technology (MOST) under its Vanguard Program 106-2119-M-002-011.

\bibliographystyle{Styles/JHEP}
\renewcommand{\bibname}{References}
\bibliography{references.bib} 

\appendix

\section{Obtaining and using \icemc}
\icemc can be obtained from the GitHub repository \url{https://github.com/anitaNeutrino/icemc}.
The only mandatory requirement for the compilation of \icemc is a ROOT~\cite{brun1997root} installation.
The compilation can be executed via the Makefile or using the CMake structure.

To run \icemc one needs to define two environment variables and add them to the global path:
\begin{itemize}
    \item  \texttt{ICEMC\_SRC\_DIR} should point to the directory where the source code is saved;
    \item  \texttt{ICEMC\_BUILD\_DIR} should point to the directory where the executable programs are.
\end{itemize}

To run \icemc one can simply do:
\begin{verbatim}
./icemc -i {inputFile} -o {outputDirectory} -r {runNumber}
-n {numberOfNeutrinos} -t {triggerThreshold} -e {energyExponent}
\end{verbatim}

All of the parameters are optional and if they are not specified inputs from \texttt{inputs.conf} are used. 
The two standard input files for the ANITA-III and ANITA-IV flights come with the package.

The output directory contains a series of root files with  information about all the neutrinos simulated, as well as a text file containing the neutrino survival efficiency at different stages of \icemc, and the volumetric acceptance.

Other programs test a portion of the full \icemc program:
\begin{itemize}
    \item {\tt testThermalNoise} simulates only the thermal noise at the payload for a specific ANITA flight.
    \item {\tt testInputAfterAntenna} simulates the injection of an RF impulse after the antenna feed; this program is used to produce trigger efficiency scans similar to the ones taken before each ANITA flight.
    \item {\tt testWAIS} simulates the WAIS pulser as described in Subsection~\ref{subsec:wais}.
\end{itemize}

To produce ANITA-like output files and use more advanced features of \icemc, the installation of \texttt{libRootFFTWwrapper} (\url{https://github.com/nichol77/libRootFftwWrapper/}) and the ANITA \texttt{eventReaderRoot} (\url{https://github.com/anitaNeutrino/eventReaderRoot}) is necessary.

\end{document}